\documentclass[aps,prb,floatfix,amsmath,amssymb,column,preprint,eqsecnum]{revtex4}
\usepackage{graphicx}
\usepackage{dcolumn}
\usepackage{bm}
\usepackage{setspace}   
\newcommand{\beq}{\begin{equation}}
\newcommand{\eeq}{\end{equation}}
\newcommand{\ba}{\begin{array}{ccc}}
\newcommand{\ea}{\end{array}}
\newcommand{\nn}{\nonumber}
 \renewcommand{\d}{\partial}
\def\bea{\begin{eqnarray}}
\def\eea{\end{eqnarray}}

\def\psiu{\psi_{\uparrow}}
\def\psid{\psi_{\downarrow}}
\def\psiud{\psi^{\dagger}_{\uparrow}}
\def\psidd{\psi^{\dagger}_{\downarrow}}

\def\<{\langle}
\def\>{\rangle}
\usepackage{amsmath}
\usepackage{amssymb}

\newcommand{\NCCO}{Nd$_{2-x}$Ce$_{x}$CuO$_{4-y}$~}
\newcommand{\PCCO}{Pr$_{2-x}$Ce$_{x}$CuO$_{4-y}$~}

\preprint{arXiv:0804.1794}

\begin{document}
\title{Destruction of N\'eel order in the cuprates by electron doping}

\author{Ribhu K. Kaul}
\affiliation{Department of Physics, Harvard University, Cambridge MA
02138}

\author{Max A. Metlitski}
\affiliation{Department of Physics, Harvard University, Cambridge MA
02138}

\author{Subir Sachdev}
\affiliation{Department of Physics, Harvard University, Cambridge MA
02138}

\author{Cenke Xu}
\affiliation{Department of Physics, Harvard University, Cambridge MA
02138}

\date{June 11, 2008\\
\vspace{1.6in}}
\begin{abstract}
Motivated by the
evidence~\cite{greene04,greene07a,greene07b,greene07c,greven07,dai07} in
\PCCO and \NCCO of a magnetic quantum critical point at which N\'eel order
is destroyed,  we study the evolution with doping of the $T=0$ quantum
phases of the electron doped cuprates.
At low doping, there is a metallic N\'eel state with small electron Fermi
pockets, and this yields a fully gapped
$d_{x^2-y^2}$ superconductor with co-existing N\'eel order at low
temperatures. We analyze the routes by which the spin-rotation symmetry
can be restored in these metallic and superconducting states.
In the metal, the loss of N\'eel order leads to a topologically ordered
`doublon metal' across a deconfined critical point with global O(4)
symmetry. In the superconductor, in addition to the conventional
spin density wave transition, we find a variety of unconventional possibilities, including
transitions to a nematic superconductor and to valence bond supersolids.
Measurements of the spin correlation length and of the anomalous dimension
of the N\'eel order by neutron scattering or NMR
should discriminate these unconventional transitions from spin
density wave theory.
\end{abstract}

\maketitle

\section{Introduction}
\label{sec:intro}

\begin{figure}
  \includegraphics[width=3in]{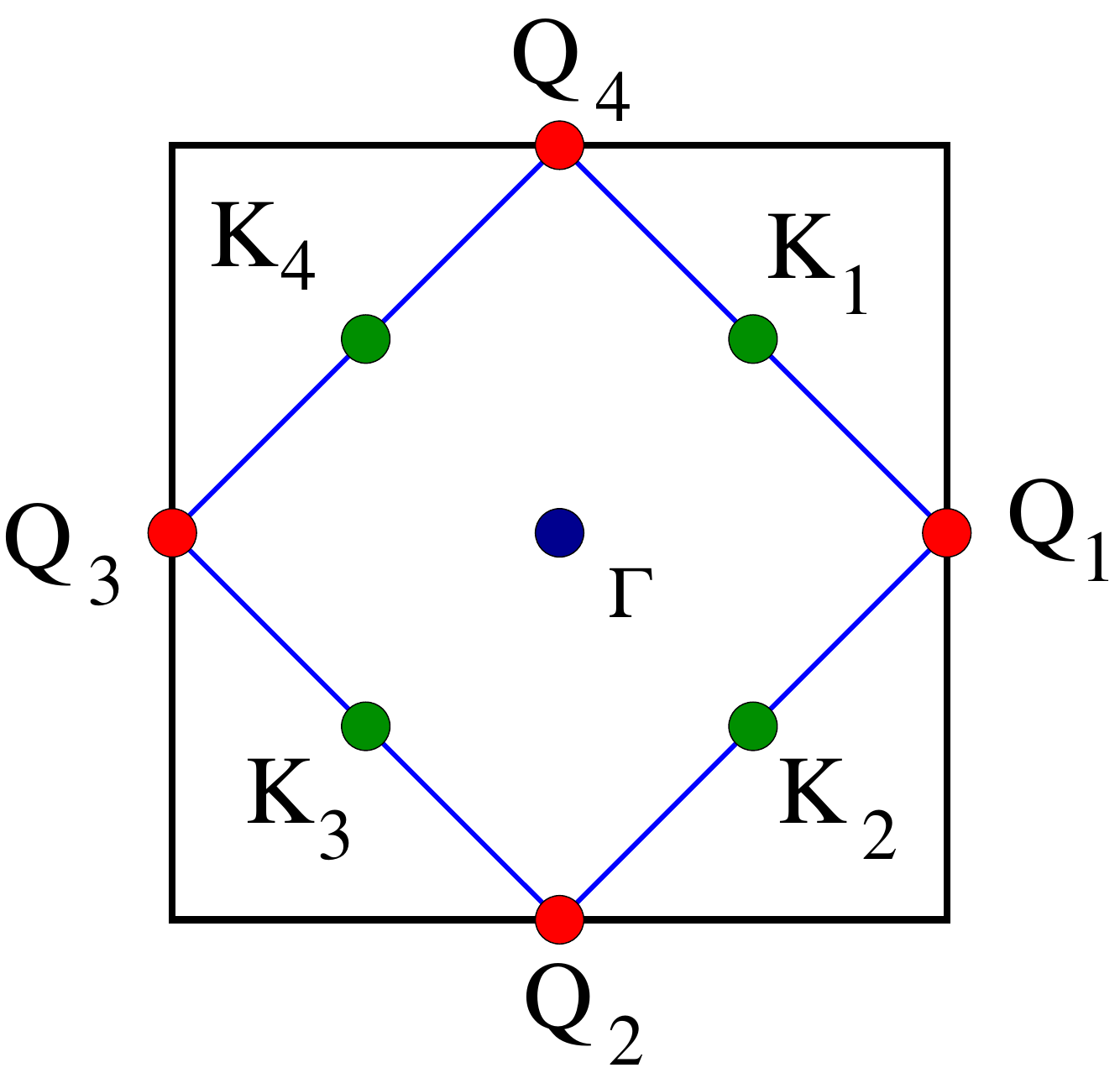}
  \caption{Brillouin zone map showing where the low energy fermions reside for hole-doped [$K_v$] and electron-doped [$Q_v$] cuprates, as deduced from photo-emission data at low doping. }
  \label{fig:bz}
\end{figure}

Superconductivity in the cuprates emerges on doping an antiferromagnetic insulator with either holes or electrons. The hole-doped cuprates generally have higher superconducting critical temperatures, but at the same time display a host of complicated phenomena, {\em e.g.\/} incommensurate magnetism and charge order, especially in the La series of compounds. 
The electron-doped cuprates on the other hand, provide an interesting contrast, where the phenomenology appears to be relatively simple.
The superconductivity also has $d$-wave pairing \cite{orderphase}, but there is no evidence yet for charge order, and the magnetic correlation remain commensurate even after long-range magnetic order is destroyed. The sharp contrast between electron and hole doping must arise from particle-hole asymmetry in Cu-O planes. The electron-hole asymmetry of the Cu-O plane is evidenced most clearly by photo-emission experiments~\cite{armitage03,claesson04,matsui05,park07} that show a sharp distinction between the Brillouin zone location of the low-energy fermions in the very lightly hole- [$K_v=(\pm \pi/2,\pm\pi/2)$] and electron- [$Q_v=(\pi,0),(0,\pi)$] doped cuprates, see Fig.~\ref{fig:bz}.

A further motivation for the study of the electron-doped cuprates is provided by recent quantum oscillation evidence for
the presence of electron pockets in the hole-doped cuprates in a strong magnetic field \cite{louis}. It seems most natural to us that
these electron pockets reside near the $Q_v$. So it seems appropriate to study the physics of the electron pockets where they are already
present in zero field: in the electron-doped cuprates. Conversely, as we will see in this paper, the hole pockets near the $K_v$ also
play a role in the physics of the electron-doped cuprates. Indeed, in both the electron- and hole-doped cuprates, a central
problem is understanding how the $K_v$ hole pockets and the $Q_v$ electron pockets reconnect to form a large
Fermi surface state after the loss of magnetic order.

A recent neutron scattering study of the N\'eel correlation length~\cite{greven07} in  \NCCO~provides evidence for a quantum critical point at $x\approx 0.13$, after which the N\'eel correlation length is finite.
Remarkably, even at the optimal doping $x\approx 0.15$ (at which long range N\'eel order is lost) a large N\'eel correlation length is measured; additionally, there is no evidence for incommensurate magnetic order over the entire doping range. The relative stability of the commensurate magnetism in the electron doped cuprates should be contrasted to the La series of the hole-doped cuprates. In the latter, long range magnetic order transforms from the $(\pi,\pi)$ N\'eel vector to incommensurate ordering vectors before being destroyed at dopings typically three times smaller than in the electron-doped cuprates. 

\begin{figure}
  \includegraphics[width=4in]{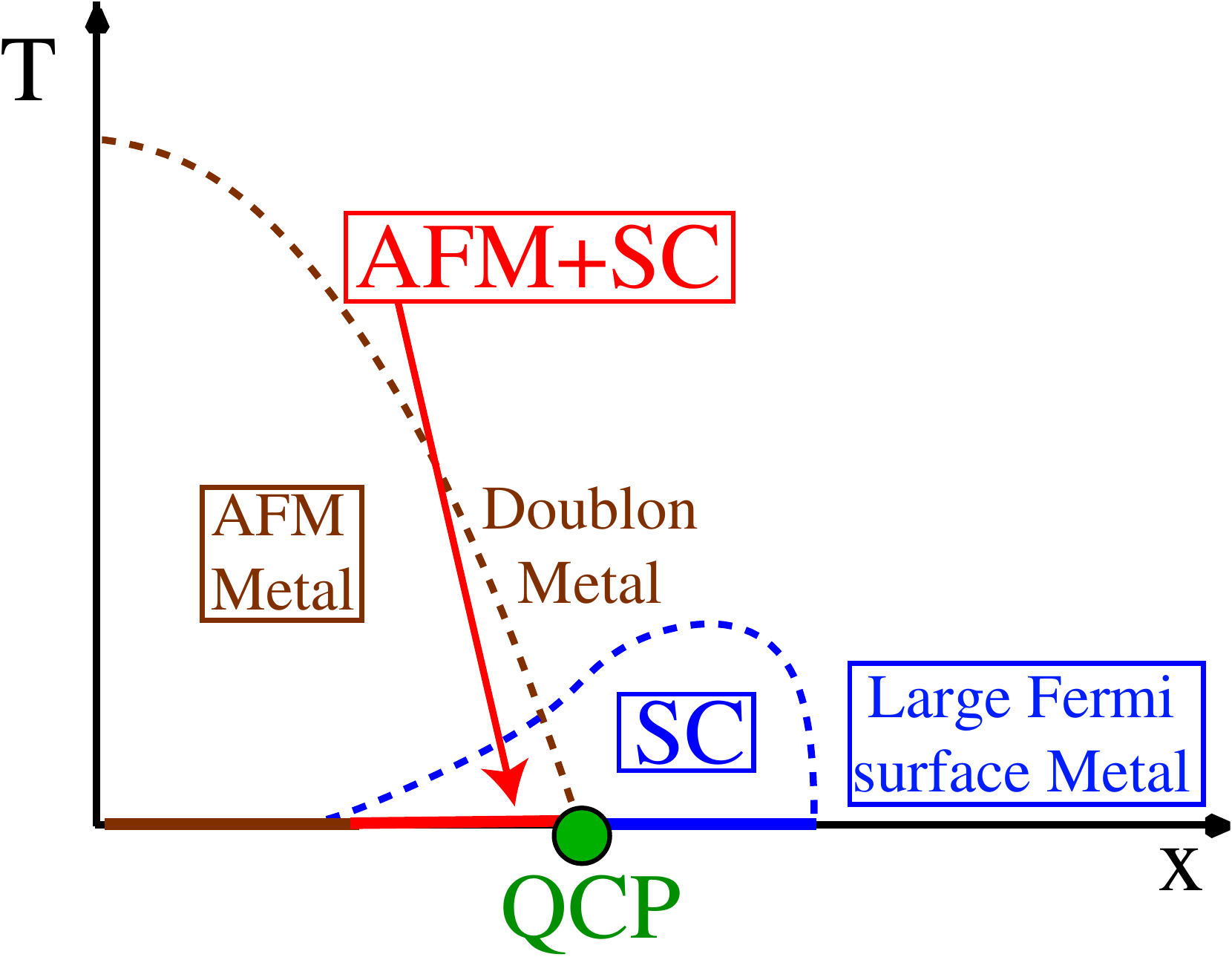}
  \caption{Schematic phase diagram for the electron doped phase cuprates (following Refs.~\onlinecite{greene04,greene07a,greene07b,greene07c,greven07,dai07}). The dashed lines indicate finite-$T$ phase transitions. The quantum critical point where N\'eel (AFM) order is lost in the superconductor (SC)
  is marked with a solid circle. The ``doublon metal'' is a phase proposed in the present paper, which appears
  when N\'eel order is lost in the AFM metal; a AFM metal/Doublon metal
  quantum critical point does not appear in the phase diagram above, but would be revealed when superconductivity is suppressed
  {\em e.g.\/} by an applied magnetic field. The finite $T$ crossovers can exhibit features of both
  the AFM+SC/SC and AFM Metal/Doublon metal quantum critical points. }
  \label{fig:Tx}
\end{figure}

These photoemission and neutron scattering measurements suggest the schematic phase diagram shown in Fig.~\ref{fig:Tx}, as a function
of temperature ($T$) and electron doping ($x$).

\begin{figure}
   \includegraphics[width=5in]{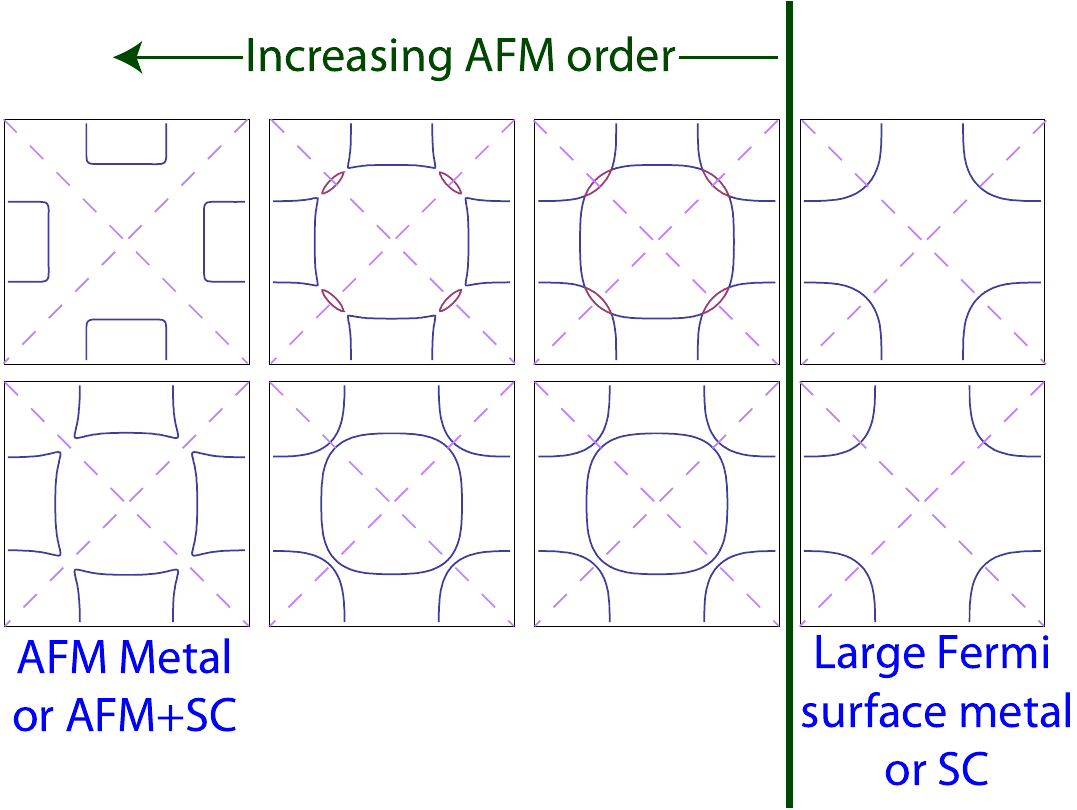}
   \caption{(color online) Fermi-surface reconstruction at SDW transition showing the presence of an intermediate state between the large Fermi surface (rightmost panels) and small Fermi-pocket (leftmost panels) states. Following Ref.~\onlinecite{linmillis}, we have used a band structure appropriate to the cuprates, $t_1=1, t_2=0.32 t_1$ and $t_3=0.5t_2$. The right most plots shows the Fermi surface before the introduction of a mean-field SDW order parameter. The second from right show the Fermi-surface after folding with $\Delta_{\rm SDW}=0$, followed by $\Delta_{\rm SDW}=0.05,0.4t_1$ moving left. The top-row has chemical potential $\mu=0.94$, and the bottom row $\mu=0.34$.
   The dashed lines indicate the points where the $d$-wave pairing amplitude changes sign in the superconducting state.
   The AFM+SC states in the leftmost panels have fully gapped quasiparticles because the Fermi surfaces do not intersect the dashed lines.
   Similarly, the large Fermi surface SC in the rightmost panels has gapless quasiparticle excitations at 4 nodal points, while the intermediate
   states have 8 nodal points.
   }
   \label{fig:fs}
 \end{figure}

The focus of this paper is on the nature of the dynamic spin correlations in the electron-doped cuprates as a function
of increasing doping. It is useful to frame our discussion by first recalling the predictions of a conventional
spin-density-wave (SDW) theory of the evolution of the Fermi surface as function
of electron density and the spontaneous N\'eel moment \cite{andrey,tremblay04,linmillis,yan06,das07}. We sketch the results
of a mean-field computation in Fig.~\ref{fig:fs}.

At very low electron doping ($x$), we have the electron Fermi-pocket states shown in the leftmost panels (AFM Metal), with well established N\'eel order.
When this state goes superconducting at low temperature ($T$),
the Fermi surface does not intersect the diagonals along which the $d_{x^2-y^2}$
pairing amplitude vanishes, and so the resulting $d$-wave superconductor (AFM+SC) is fully gapped.
At large electron doping, 
we have the large Fermi surfaces shown in the rightmost panels, with no N\'eel order. Now the Fermi surfaces
do intersect the diagonals at 4 points, and so the $d$-wave superconductor has 4 nodal points.
Examining the evolution of the Fermi surfaces between these two limiting cases in Fig.~\ref{fig:fs}, we observe that there
is generically an intermediate Fermi surface configuration, with N\'eel order, in which the Fermi surfaces intersect the diagonals at the
8 points $\pm ( \pi/2, \pi/2) \pm (\epsilon, \epsilon)$ and $\pm ( \pi/2, -\pi/2) \pm (\epsilon, -\epsilon)$, for some
small non-zero $\epsilon$. The appearance of superconductivity at low $T$ will then lead to a $d$-wave superconductor
with 8 nodal points in the full Brillouin zone of the square lattice. Thus, in both the metallic and superconducting cases,
this intermediate state has 8 zero-energy crossings of the fermion dispersion relation along the diagonals of the full
square lattice Brillouin zone.

A further motivation for our study is that the 8 diagonal Fermi points of the intermediate state are not clearly seen in photoemission experiments \cite{armitage03,claesson04,matsui05,park07}. Fermi surface crossings are seen on only a single point adjacent to the
4 $(\pm \pi/2, \pm \pi/2)$ points. We therefore explore here unconventional routes
by which the N\'eel order at low doping, in the AFM metal and the fully gapped $d$-wave superconductor, can be 
destroyed by increasing hole concentration.

\begin{figure}
   \includegraphics[width=4.5in]{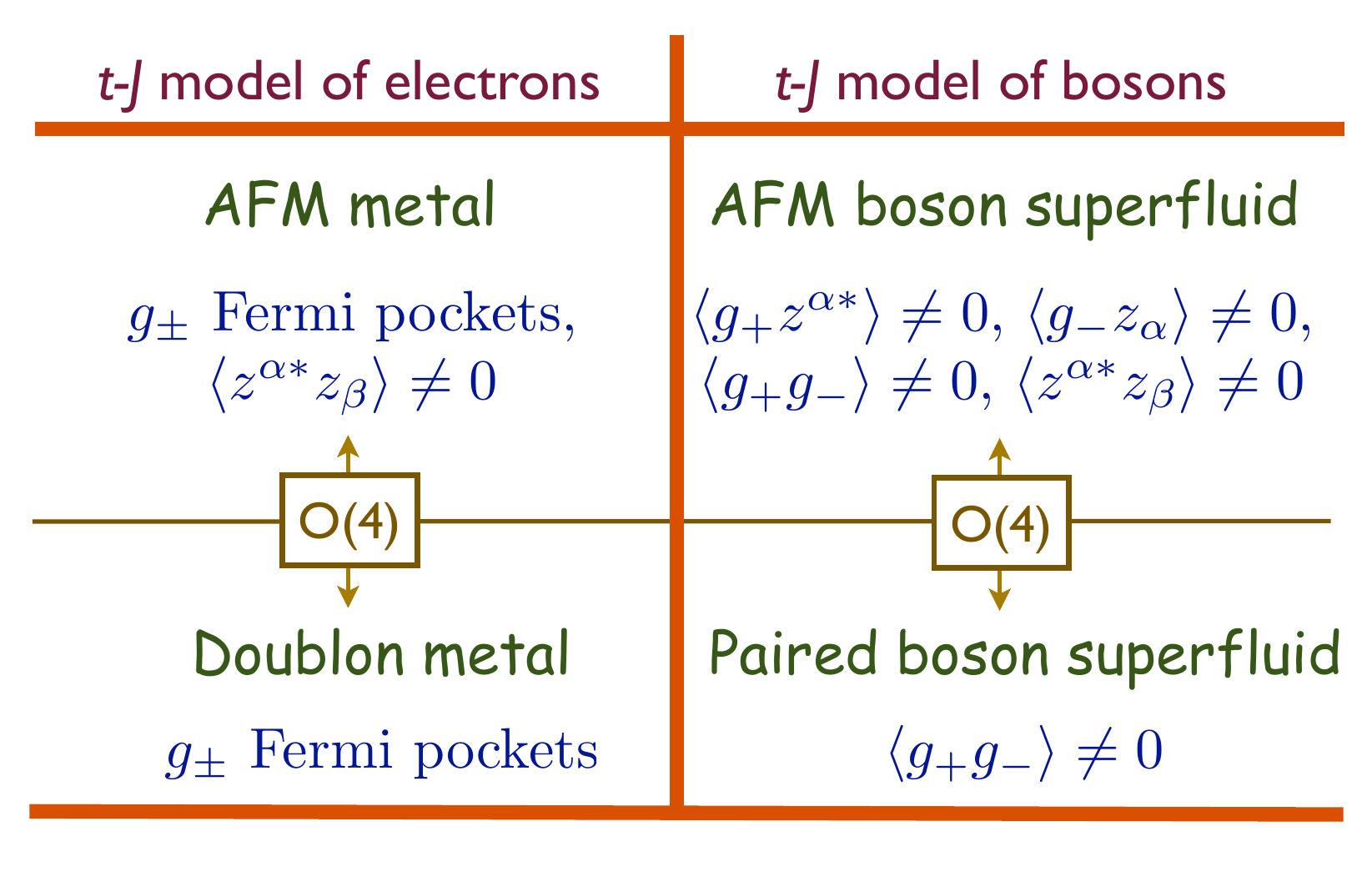}
   \caption{(color online) Analogy between the phases and phase transitions of the physical $t$-$J$ model of electrons
   and the toy $t$-$J$ model of $S=1/2$ bosons. This figure lists only the metallic phases of the electrons, considered in Section~\ref{sec:metal}.
   All non-zero, gauge-invariant condensates (bilinear in the $g_\pm$ and $z_{\alpha}$)
   of each phase are noted; those not shown are zero in that phase.
   The boson-analog of the fermionic metallic states are obtained if we replace the $g_\pm$ Fermi pockets by
   condensates of the $g_\pm$ bosons: then the ``fermionic Higgs mechanism'' discussed in Section~\ref{sec:metal} finds
   its analog in the ordinary Higgs condensate of the $g_\pm$ bosons. Monopoles are suppressed in all phases above.
  }
   \label{fig:phases1}
 \end{figure}

\begin{figure}
   \includegraphics[width=6in]{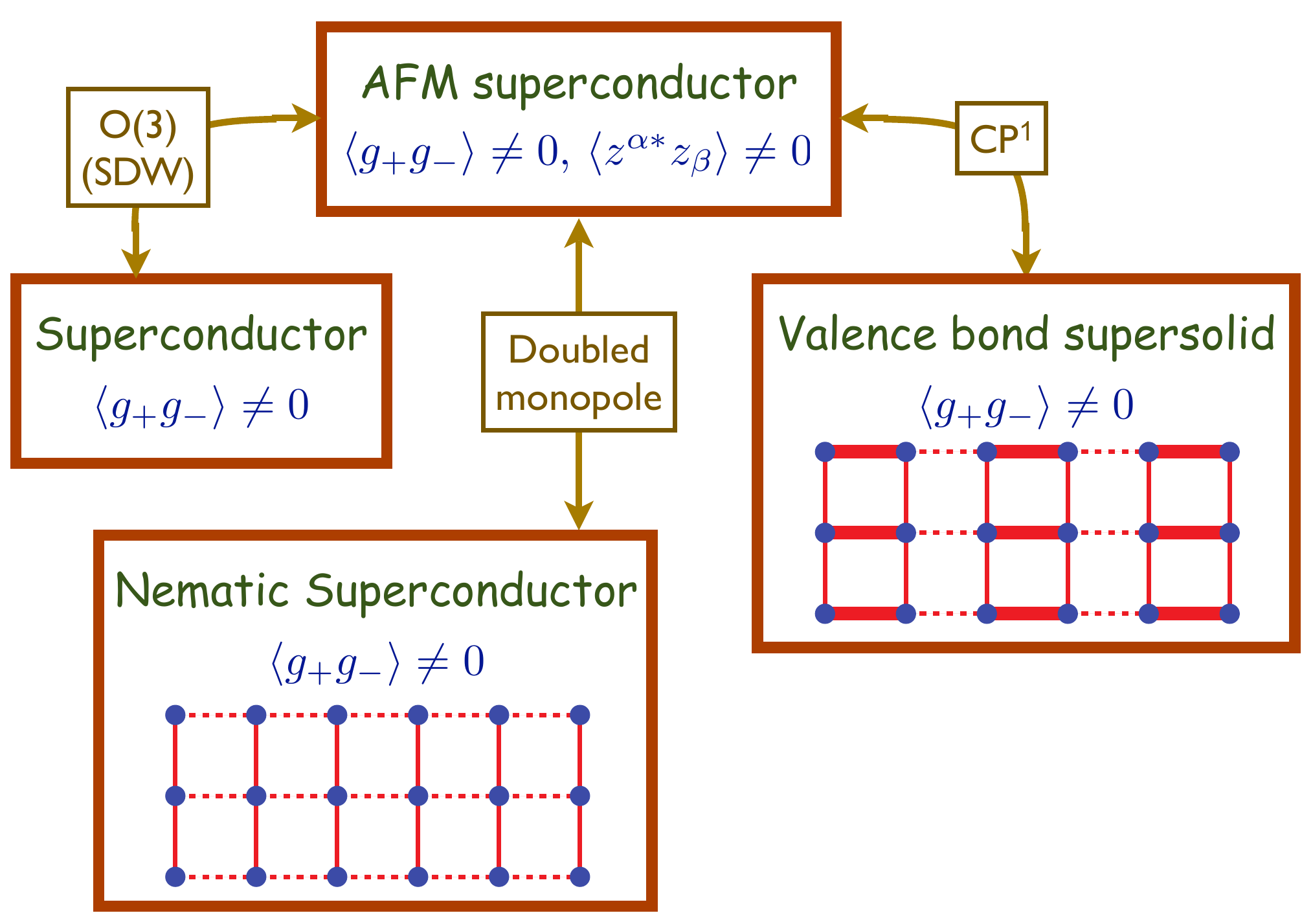}
   \caption{(color online) As in Fig.~\ref{fig:phases1}, but for the superconducting phases of the electrons discussed in Section~\ref{sec:supercond}. 
   The explicit computations in Section~\ref{sec:bosons} are for the boson model, but for paired superfluids (which includes all phases above)
   the results are expected to also apply to electrons. The pattern of translation symmetry breaking in the valence bond
   supersolid is sketched (see also Fig.~\ref{fig:ins}), while that in the nematic superconductor follows Fig.~\ref{q2}a.
   Additional forms of translational symmetry
   breaking are also possible in the AFM and non-magnetic phases, 
   and these are discussed in Section~\ref{sec:max}. 
  }
   \label{fig:phases2}
 \end{figure}

Important aspects of our results on the metallic and superconducting
quantum phases and phase transitions are summarized in
Figs~\ref{fig:phases1} and \ref{fig:phases2}.
The right panel of Fig.~\ref{fig:phases1} indicates results on a ``toy'' $t$-$J$ model of $S=1/2$ bosons
which we will describe in Section~\ref{sec:bosons}. We will see that there is a close analogy between our results
for the electronic $t$-$J$ model and the toy boson model, with the latter model having the advantage that
duality computations of the crossover into confinement can be carried out in explicit detail.

For the metallic case, we find that the quantum transition out of the N\'eel state with Fermi-pockets (the AFM Metal) is into
an exotic `doublon metal' state without magnetic order (see Fig.~\ref{fig:phases1}); the nomenclature refers to the 
sites with double occupancy when the Mott insulator is doped with electrons. The `doublon metal' is the particle-hole conjugate of the
`holon metal' state described in recent work \cite{acl,qi}, and both are examples of `algebraic charge liquids'.
These states have topological order and no sharp electron-like quasiparticles. However, they are separated
from conventional Fermi liquid states by sharp transitions only at $T=0$; at $T>0$ there are
only crossovers into the Fermi liquid-like regime. As superconductivity always appears as $T \rightarrow 0$ (see Fig.~\ref{fig:Tx}), it is these
$T>0$ crossovers of the metallic regime which are needed for experimental comparisons. We shall show that the spin
excitations near the transition into the doublon metal are described by a quantum field theory with global O(4) symmetry, as indicated
in Fig.~\ref{fig:phases1}.
Further, as we discuss below in Section~\ref{sec:expts}, spin fluctuations of this O(4) theory have clear experimental signatures.
Section~\ref{sec:bosons} will show that these metallic phases of the electronic $t$-$J$ model also have strikingly
similar analogs in the $t$-$J$ model of bosons, along with a magnetic ordering transition in the O(4) class.

For the superconducting case, we find a number of distinct possibilities, which are illustrated in Fig.~\ref{fig:phases2}. From
the AFM superconductor we find 3 distinct classes of transitions:\\
({\em a\/}) A transition to a $d$-wave superconductor with full square lattice symmetry, which is in the
O(3) universality class. This transition is in the same universality class as conventional SDW theory. It is quite remarkable, and novel,
that this SDW transition reappears in our formalism based on fractionalized degrees of freedom.\\
({\em b\/}) A transition to a $d$-wave superconductor with co-existing valence bond solid (VBS) order, {\em i.e.\/}
to a supersolid. Such a valence bond supersolid was initially discussed in Refs.~\onlinecite{rsi,vs}. The pattern of the VBS order is columnar or plaquette (see Figs.~\ref{fig:phases2} and \ref{fig:ins}), 
the same as that in the insulator \cite{ssnature,rsb}
(for rational $x/2 = p/q$ with $q/2$ odd, other patterns of order are possible, as discussed in Section~\ref{sec:gammapp}).
This transition is expected to be of the `deconfined' variety, in the CP$^{1}$ universality class, similar to the transition in insulating antiferromagnets \cite{senthil1,ssnature}.\\
({\em c\/}) The third transition is described by the CP$^{1}$ theory, but with an additional `doubled monopole' perturbation allowed,
which will be explained in more detail in Sections~\ref{sec:max}. The non-magnetic superconductor does break the 
square lattice space group symmetry, and the two allowed patterns of symmetry breaking are in Fig.~\ref{q2}. Notice that
one them only breaks the $Z_4$ rotation symmetry of the square lattice to a $Z_2$ rectangular symmetry, leading to
a {\em nematic superconductor\/} shown in Fig.~\ref{fig:phases2}.

It is interesting to note that the above possibilities match the patterns of transitions found \cite{rsb} for insulating
antiferromagnets as function of the spin $S$. In particular, case ({\em a\/}) occurs for even integer $S$, 
case ({\em b\/}) for half-odd-integer $S$, and case ({\em c\/}) for odd integer $S$. Here we are considering a $S=1/2$ antiferromagnet,
but with a background of a compressible superconductor. As we shall see, the background density fluctuations
in the superconductor are able to modify the spin Berry phases so that the transitions match those for different $S$ in insulators.

In addition to cases ({\em a\/}), ({\em b\/}), ({\em c\/}), we note briefly that 
it is also possible that the AFM+SC state already has density modulations.
Then, the transition involving loss of AFM order will lead to modifications in the ordering pattern, as will
be discussed in some detail in Section~\ref{sec:max}. An important point is that, in all these cases, the set of allowed periods for the density modulations in the supersolid without AFM order are the same as those characteristic \cite{bbbss1} of paired
supersolids of density $1+x$. 

An interesting issue, which we shall largely leave open in the present paper, is the nature of the spectrum
of the fermionic Bogoliubov quasiparticle excitations in the supersolid or the nematic superconductor. 
One natural possibility \cite{vojtass,granath,parkss} is that these
excitations initially remain fully gapped, as in the AFM+SC state. 
On the other hand, knowing that the supersolid or nematic superconductor has
effective density $1+x$, the structure of the Fermi surfaces in Fig.~\ref{fig:fs} suggest that such a $d$-wave superconductor should have 4 gapless Dirac points. 
In the deconfined quantum critical theory, the electron spectral function is fully gapped along the diagonals of the Brillouin zone.
If the gapless nodal points do appear in the non-magnetic phase, 
they would create ``Fermi liquid coherence peaks'' at the nodal points, with the
weight of the coherence peak vanishing as we approach the quantum critical point. This phenomenon would then
resemble that in dynamical mean field theory \cite{dmft,dsf}, where the Fermi liquid coherence peaks of the metal vanish at the metal-insulator
transition, revealing a fully gapped single-particle spectrum at the critical point. 
This issue will be discussed further in Section~\ref{sec:supercond}.

\subsection{Experimental tests}
\label{sec:expts}

We note here that neutron scattering or NMR measurements of the spin excitation spectrum can serve as
useful experimental probes of whether the N\'eel order is lost as in a conventional SDW framework,
or in a more exotic deconfined critical point . In particular, the temperature dependence of various
components of the dynamic structure factor in the quantum critical region can measure two crucial exponents
characterizing the transition, the dynamic critical exponent, $z$, and
the anomalous dimensions of the N\'eel order parameter, $\eta_N$.
In terms of these exponents, we have \cite{csy} for $S_N^e$, the zero frequency  dynamic structure factor at the N\'eel ordering wavevector (proportional
to the elastic neutron scattering cross-section at $(\pi, \pi)$):
\begin{equation}
S_N^e \sim T^{(-2 + \eta_N)/z};
\end{equation}
for, $S_N$, the equal-time structure factor at the N\'eel ordering wavevector (proportional
to the energy-integrated neutron scattering cross-section at $(\pi, \pi)$):
\begin{equation}
\label{eq:eqtimeS}
S_N \sim T^{(-2 + z + \eta_N)/z};
\end{equation}
for $\xi$, the N\'eel correlation length:
\begin{equation}
\xi \sim T^{-1/z};
\end{equation}
The present neutron scattering experiment \cite{greven07} only reports the quantum critical
behavior of the spin correlation length, which is consistent with $z=1$. Although data on $S_N$ exists, a scaling analysis to extract the exponent Eq.~(\ref{eq:eqtimeS}) has not been carried out. An important test of quantum critical scaling would be to check that the exponent that arises from this analysis should agree with an extraction of the same index by an analysis of the Cu NMR relaxation rate,
\begin{equation}
\label{eq:t1}
\frac{1}{T_1} \sim T^{\eta_N/z}.
\end{equation}

\begin{table}[t]
\begin{spacing}{1.5}
\centering
\begin{tabular}{||c||c|c||c|c||} \hline\hline
 & SDW-metal& SDW-SC & ~O(4)~ & ~CP$^1$~  \\
 \hline\hline
$z$ & 2  & 1 & 1 & 1
\\ \hline
$\eta_N$ & 0 & 0.038 & 1.37  & 0.35   \\
\hline\hline
\end{tabular}
\end{spacing}
\caption{Predictions for the exponents $\eta_N$ and $z$ by different theories for the quantum critical point observed in the electron-doped cuprates such as \NCCO. Both exponents can be measured in experiment by a straightforward analysis of the temperature dependence of the equal-time structure factor, as described in the text. The numerical estimates for the anomalous dimensions are based on results from previous studies of the 3-dimensional O(3) [Ref.~\onlinecite{Camp}] , 3-dimensional O(4) [Ref.~\onlinecite{vicari,kim}] and the 3-dimensional CP$^1$ model inferred from quantum simulations of the N\'eel-VBS transition [Ref.~\onlinecite{melkokaul}]. }
\label{exptable}
\end{table}

The values of the exponents in the conventional SDW theory depend upon whether the
quantum critical region is controlled by a metallic or a superconducting fixed point.
For the metallic fixed point, we have the Hertz-Millis-Moriya theory~\cite{hertzmillis} $z=2$ and $\eta_N=0$, while for the superconducting
case we have the usual 3D O(3) transition, $z=1$ and $\eta_N \approx 0.038$.

Our main new experimentally relevant results in this paper are the values of these exponents for the `deconfined' transition
 at which N\'eel order is lost.
The exponents depend upon whether we are using a superconducting or metallic fixed point,
and our results are summarized in Table~\ref{exptable}. There are no existing numerical results for the `doubled monopole'
transition, and so these are not shown: it may well be that this case has a first-order transition.
Note the large values of $\eta_N$ for the deconfined cases, making them clearly distinguishable from the SDW cases.
In particular, with $\eta_N > 1$ for the metallic case, the equal-time structure factor, $S_N$
has a singular contribution which {\em decreases\/} with decreasing $T$.

We also note that for the superconducting case, the properties of the CP$^1$ field theory are not
fully settled in the literature \cite{melkokaul,kuklov,shailesh} with a debate on whether the quantum transition
is second- or first-order. Nevertheless, there is significant evidence \cite{sandvik,ssnature} of a crossover
into a regime which is described by the CP$^1$ field theory. Furthermore, even if the transition is first-order,
it appears to be only very weakly so, and the simulations of Ref.~\onlinecite{melkokaul} show a substantial
$T>0$ critical scaling regime.

Because the electron-doped cuprates are always superconducting in the proximity of the quantum critical point at low $T$, the superconducting critical theory described above is the correct description at very low-$T$ scales.  The normal state theory does however apply at temperature scales above the superconducting temperature and hence could be the relevant one for experiments over a large temperature scale. An interesting prediction that arises from this crossover is that the equal-time structure factor, $S_N$, could have a non-monotonic $T$ dependence. It should first decrease with cooling (when the system is controlled by the metallic fixed point with $\eta_N >1$), and then crossover to increasing with further cooling, when the system is controlled by the superconducting fixed point with $\eta_N < 1$.

The outline of the remainder of this paper is as follows: In Sec.~\ref{sec:fields} we derive an effective field theory for the electron-doped cuprates in a language well suited to discuss both the magnetic phases and the non-magnetic ones that appear on
the destruction of N\'eel order. In Sec.~\ref{sec:qc} we discuss the various possibilities for transitions involving loss
of N\'eel order.
The $t$-$J$ model of bosons will be introduced in Section~\ref{sec:bosons}, along with a complete duality analysis of its
phase diagram and its crossover to confining phases.
Finally in Sec.~\ref{sec:conc}, we conclude with a summary of our results.

\section{Field Theory at Low Doping}
\label{sec:fields}

We begin with a symmetry-based derivation of a long wavelength effective action for the electron-doped cuprates.
We will use the low energy excitations of the low doping state to build a theory which is valid
also at larger doping when spin rotation invariance is restored.

The motion of a small number of charge carriers in a quantum anti-ferromagnet is usually described by the $t-J$ model,
\begin{equation}
\label{tJmodel} H_{t-J} = -
\sum_{i,j,\alpha}t_{ij}(c^{\alpha\dagger}_{i}c_{j\alpha} + {\rm
h.c.}) + \sum_{i,j} J_{ij}\vec{S}_i \cdot \vec{S}_j + \ldots \, ,
\end{equation}
where $c_{i\alpha}$ destroys an electron with spin $\alpha$
on the sites $i$ of a square lattice and
$\vec{S}_i=\frac{1}{2}\sum_{\alpha\beta} c^{\alpha\dagger}_{i}
\vec{\sigma}_{\alpha}^{\beta}c_{i\beta}$, with $\vec{\sigma}$ the
Pauli matrices. We shall study the case in which the electrons hop on a square lattice.
Once extra electrons are doped into the half-filled magnet
a constraint must also be included. The constraint,
\begin{eqnarray}
\label{eq:elcons}\sum_{\alpha}c^{\alpha\dagger}_{i}c_{i\alpha} &\geq& 1 
\end{eqnarray}
is enforced on
each site, modeling the large local repulsion between the electrons.
It is important to note that our results are more general than a
particular $t$-$J$ model, and follow almost completely from symmetry
considerations. The ellipses in Eq.~(\ref{tJmodel}) indicate additional
short-range couplings which preserve square lattice symmetry and
spin rotation invariance.

\begin{table}[!t]
\begin{spacing}{2}
\centering
{\em LATTICE FIELDS:}\\[5 pt]
\begin{tabular}{||c||c|c|c|c||} \hline\hline
 & $T_x$ & $R_{\pi/2}^{\rm dual}$ & $I_x^{\rm dual}$ & $\mathcal{T}$  \\
 \hline\hline
$~~b_\alpha$~~ & $ ~\varepsilon_{\alpha\beta} \overline{b}^{\beta}~$
& $ ~\varepsilon_{\alpha\beta} \overline{b}^{\beta }~$ & $
~\varepsilon_{\alpha\beta} \overline{b}^{\beta }~$ & $
~\varepsilon_{\alpha\beta} b^{\beta \dagger}~$\\
\hline
$\overline{b}^\alpha$ & $ \varepsilon^{\alpha\beta}
{b}_{\beta}$ & $ \varepsilon^{\alpha\beta} {b}_{\beta }$ & $
\varepsilon^{\alpha\beta} {b}_{\beta }$ & $
\varepsilon^{\alpha\beta} \overline{b}_{\beta}^{\dagger}$\\
\hline
$g_{+}$ & $ g_{-}$ & $ g_{-}$ & $ g_{-}$  &  $-g_{+}^\dagger $ \\
\hline
$g_{-}$ & $-g_{+}$ & $ -g_{+}$ & $ -g_{+}$  & $-g_{-}^\dagger$ \\

\hline\hline
\end{tabular}
\end{spacing}
\caption{Transformations of the lattice fields under square
lattice symmetry operations and time reversal. $T_x$: translation by one lattice
spacing along the $x$ direction; $R_{\pi/2}^{\rm dual}$: 90$^\circ$
rotation about a dual lattice site on the plaquette center
($x\rightarrow y,y\rightarrow-x$); $I_x^{\rm dual}$: reflection
about the dual lattice $y$ axis ($x\rightarrow -x,y\rightarrow y$);
$\mathcal{T}$: time-reversal, defined as a symmetry of the imaginary
time path integral. The transformations of the Hermitian conjugates
are the conjugates of the above, except for time-reversal of
fermions. For the latter, $g_{\pm}$ and $g^\dagger_{\pm}$ are treated as independent Grassman numbers and
$\mathcal{T}: g^\dagger_{\pm} \rightarrow  g_{\pm}$.}
\label{table0}
\end{table}

Following Ref.~\onlinecite{ffliq}, but now for the case of electron-doping, we re-write the electron operators in a $t-J$ type lattice model in terms of spinons and `doublons' (for doubly occupied sites). 
Note that here the site occupation is constrained to be $\sum_{\alpha}c^\dagger_{\alpha}c_\alpha \geq 1$. We use the following representation for the electron operators,
\begin{eqnarray}
c_{\alpha}&=& \varepsilon_{\alpha\beta}b^{\beta\dagger} g_+ \mbox{ (on A)}\nonumber \\
c_{\alpha}&=&- \overline{b}^\dagger_\alpha g_- \mbox{ (on B)}
\label{eq:constraint}
\end{eqnarray}
where the constraint is $b^{\alpha \dagger}b_\alpha+ g^\dagger_+g_+ =1$. [We first used $c_\alpha= \varepsilon_{\alpha\beta}b^{\beta\dagger}g$ on both sub-lattices, then rotated the Schwinger bosons on the B sublattice $b_\alpha \rightarrow \varepsilon_{\alpha\beta}\overline{b}^\alpha$, like in the Auerbach-Arovas analysis~\cite{aa}]. Note: $\varepsilon^{\alpha\beta}\varepsilon_{\beta\gamma}=-\delta^\alpha_\gamma$

Now we are in a position to write down the transformation of the lattice fields that we have written down under the various square-lattice symmtries and time reversal. We require that the composite fields $c_{\alpha}$ transform into each other in the usual way under the square lattice symmetries. The implementation of time reversal symmetry is detailed in Appendix~\ref{app:trs}. We thus arrive at the Table~\ref{table0}.


We now proceed to take the continuum limit of the lattice model that we have defined. In order to do so~\cite{rsb}, we define fields $z_\alpha = b_\alpha+ \overline{b}_\alpha^\dagger$ and $\pi_\alpha=b_\alpha-\overline{b}_\alpha^\dagger$ and integrate out the massive $\pi_\alpha$ field. We then arrive at the Lagrangian for the $z_\alpha$
\begin{equation}
\mathcal{L}_{z} =D^+_\mu z^{\alpha *}    D^-_\mu z_\alpha
+ s |z_\alpha|^2 + u \left( |z_\alpha |^2 \right)^2 + \ldots
\end{equation}
where $\mu = x,y, \tau$ is a spacetime index,
$D^\pm_\mu=\partial_\mu \pm iA_\mu$, $A_\mu$ is an emergent U(1) gauge field linked to the local
constraint in Eq.~(\ref{eq:constraint}), and $s$ and $u$ are couplings which can be tuned to explore the phase diagram.
The N\'eel order parameter is simply $\vec{n}=z^{\alpha*} \vec{\sigma}^\beta_{\alpha} z_\beta$.

We also need to take the continuum limit for the charge carrying fermions of this model. As discussed in detail in Ref.~\onlinecite{ffliq}, fermions that live on opposite sub-lattices carry opposite charges under the gauge field, $A_\mu$, and hence must be represented by two distinct continuum fields $g_\pm$ (both fields are centered at the lattice momentum, $Q_1$). The lowest derivative term consistent with the symmetry of the $g_\pm$ is,
\begin{equation}
\label{eq:Lg} \mathcal{L}_g =  \sum_{q=\pm}
g^\dagger_{q}  \biggl(
D^{\overline{q}}_\tau -\mu - \frac{ D_j^{\overline{q}2}}{2m} \biggr) g_{q}.
\end{equation}
where $m$ is the curvature of the fermion bands and $\overline{q}=-q$. Finally, by requiring consistency with the lattice transformation properties of the continuum fields, presented in Table~\ref{psgtable}, the lowest allowed derivative term that couples the opposite fermions $g_\pm$ can be deduced,
\begin{eqnarray}
\mathcal{L}_{z-g}&=&
\lambda ~\varepsilon^{\alpha\beta}\left[g_+^\dagger  \left( D^+_x g_- \right) z_\alpha \left(D^-_x z_\beta \right)
-g_+^\dagger \left( D^+_y g_- \right) z_\alpha \left(D^-_y z_\beta \right) \right] \\
&+& \varepsilon_{\alpha\beta}\left[g_-^\dagger \left(D^-_y g_+ \right) z^{\alpha*} \left(D^+_y z^{\beta *} \right)
- g_-^\dagger \left(D^-_x g_+ \right) z^{\alpha*} \left(D^+_x z^{\beta *} \right)\right] + {\rm c.c.}\nonumber
\label{lc2}
\end{eqnarray}
\begin{table}[t]
\begin{spacing}{1.5}
\centering
{\em CONTINUUM FIELDS:}\\[5 pt]
\begin{tabular}{||c||c|c|c|c||} \hline\hline
 & $T_x$ &$R_{\pi/2}^{\rm dual}$ & $I_x^{\rm dual}$ & $\mathcal{T}$  \\
 \hline\hline
$z_\alpha$ & $ ~\varepsilon_{\alpha\beta} z^{\beta \ast}~$  & $
~\varepsilon_{\alpha\beta} z^{\beta \ast}~$ & $
~\varepsilon_{\alpha\beta} z^{\beta \ast}~$ & $
~\varepsilon_{\alpha\beta} z^{\beta \ast}~$
\\
\hline $g_{+}$ & $- g_{-}$ & $-g_{-}$ & $-g_{-}$  & $-g_{+}^\dagger$   \\
\hline $g_{-}$ & $ g_{+}$ & $- g_{+}$ & $ g_{+}$  & $-g_{-}^\dagger$ \\
\hline\hline
\end{tabular}
\end{spacing}
\caption{Transformation properties under square lattice symmetries and time reversal, of continuum fields entering the effective action. Conjugate fields transform into the conjugate of the transformed fields except for $\mathcal{T}: g^\dagger_\pm \rightarrow g_\pm$  }
\label{psgtable}
\end{table}
This is the analog for electron-doped cuprates, of the well-known Shraiman-Siggia term~\cite{ss} in the hole-doped case. Remarkably, this term has two spatial derivatives; there is no term allowed with a single spatial derivative (as is found from a similar analysis in the hole-doped case~\cite{ffliq}; see also~\cite{wiese}). The extra derivative makes the effect of this term weaker. The weakness of this coupling, which arises because of the BZ location of the low energy fermions, (which in turn is ultimately tied to the p-h asymmetry in the Cu-O layers) is the fundamental reason for the robustness of the commensurate N\'eel correlations in the electron-doped cuprates as compared to the hole-doped case. These correlations extend at least up to optimal doping~\cite{greven07,dai07} and possibly beyond giving us confidence in
the present approach.

The complete effective action for the electron-doped antiferromagnet is then
$\mathcal{S}=\int d^2r d\tau (\mathcal{L}_z+ \mathcal{L}_g + \mathcal{L}_{z-g}) + \mathcal{S}_B$.
The final term, $\mathcal{S}_B$ contains the Berry phases of the monopoles, and has the form
\begin{equation}
\mathcal{S}_B = i \frac{\pi}{2} \sum_\jmath m_\jmath \zeta_\jmath
\label{eq:sb}
\end{equation}
for monopoles with integer charges $m_\jmath$ on the sites $\jmath$ on the dual lattice; $\zeta_\jmath$ is fixed at
$\zeta_\jmath = 0,1,2,3$
on the four dual sublattices \cite{rsb}.

\subsection{N\'eel order and superconductivity}
\label{sec:coex}

We now discuss the phase diagram of the field theory presented in the previous section. Some of the analysis
parallels that presented in Refs.~\onlinecite{ffliq} and~\onlinecite{acl} for the hole-doped case.

The phases are most easily characterized by using a representation for
the physical electron annihilation operator
$\Psi_{\alpha}(\vec{r})$ in terms of
the fields we have introduced above. We first express the electron operator in terms of its components
at momenta at $Q_1$ and $Q_2$,
\begin{equation}
\Psi_{\alpha} (\vec{r}) = e^{i \vec{Q}_1 \cdot \vec{r}} \Psi_{1\alpha}(\vec{r})+  e^{i \vec{Q}_2 \cdot \vec{r}} \Psi_{2\alpha}(\vec{r}).
\label{eq:pp}
\end{equation}
Then, as in Ref.~\onlinecite{ffliq}, we use the symmetry transformation properties to deduce the unique bilinear combination of the fermion and and CP$^1$ fields that transform in the way that the
physical electrons $\Psi_{1,2}$ should,
\begin{eqnarray}
\Psi_{1,2\alpha}&=& \varepsilon_{\alpha\beta}z^{\beta*} g_{+} \mp z_{\alpha}g_{-}.
\label{eq:physel}
\end{eqnarray}

The phases is Fig.~\ref{fig:Tx} can now be characterized in terms of the $z_\alpha$ and $g_\pm$ degrees of
freedom:\\
({\em i\/}) \underline{AFM metal}: This is the Higgs phase of the gauge theory, in which there is Higgs condensate
of $z_\alpha$ with $\langle z_\alpha \rangle \neq 0$. As discussed in Ref.~\onlinecite{ffliq}, the ``Meissner'' effect
associated with this Higgs condensate ties the $A_\mu$ gauge charge to the spin quantum number. So for
N\'eel order oriented along the $z$ axis, the $g_\pm$ fermions carry spin $S_z = \pm 1/2$ and reside in Fermi
pockets. The resulting phase is then identical to the AFM metal phase obtained in SDW theory, and shown
in the left panel of Fig.~\ref{fig:fs}.\\
({\em ii\/}) \underline{Doublon metal}: This is the particle-hole conjugate of the holon metal, and is a non-Fermi liquid
`algebraic charge liquid'. We have $\langle z_\alpha \rangle =0$, and and the phase is described
by the gapped $z_\alpha$ quanta and the $g_\pm$ Fermi pockets interacting via exchange of the $A_\mu$
gauge force. We observe from Eq.~(\ref{eq:physel}) that the physical electron involves a convolution
of the propagators of the $z_\alpha$ and $g_\pm$, and so will not have Fermi liquid form.
We note that the holons and the holon-spinon bound states, discussed in previous work \cite{ffliq,acl} on the holon metal,
are also legitimate excitations of the doublon metal. Here they are likely to be gapped, but at $T>0$ will contribute
photoemission spectral weight \cite{armitage03} near the $K_v$ points in Fig~\ref{fig:bz}.
\\
({\em iii\/}) \underline{SC phases}: As discussed in Ref.~\onlinecite{acl}, the nearest-neighbor hopping term,
and the gauge forces, will induce a pairing of the $g_\pm$ fermions. Let us assume a pairing of the form
\begin{equation}
\langle g_{+1}(k)g_{-1}(-k)\rangle = \Delta(k).
\label{eq:gpair}
\end{equation}
Then the pairing signature of the electrons can be computed from Eq.~(\ref{eq:physel}) and
(\ref{eq:gpair}): the various possibilities are discussed below. If we also have $\langle z_\alpha \rangle \neq 0$,
then we obtain the AFM+SC phase of Fig.~\ref{fig:Tx}. This is a stable phase, because the $z_\alpha$ Higgs
condensate quenches the gauge fluctuations and also the monopoles; its physical properties are
identical to the AFM+SC phase obtained in the SDW theory noted in Fig.~\ref{fig:fs}. A superconducting phase
with $\langle z_\alpha \rangle = 0$ is the doublon superconductor, and this is not stable: proliferation
of monopoles will lead to confinement, as we shall discuss in Section~\ref{sec:supercond}.

\begin{figure}
   \includegraphics[width=3in]{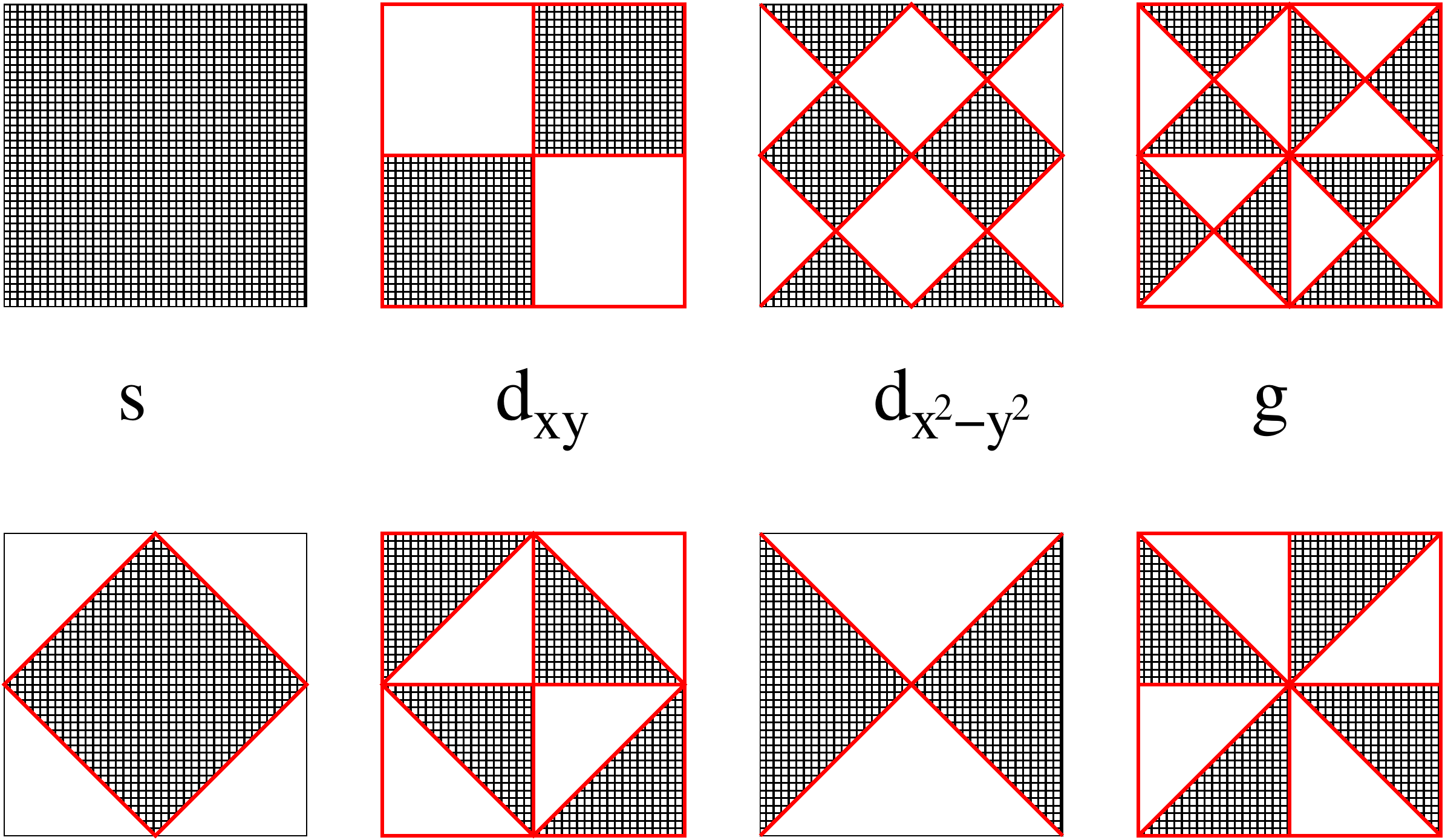}
   \caption{(color online) Various order parameters that are even under inversion $\Delta(k)=\Delta(-k)$. Shaded areas are positive and white areas negative; Thick red lines denote  zeros of the order parameter. The first (second) row has order parameters that are even (odd) under translation by a N\'eel vector $\Delta(k+K_N)=-\Delta(k)$. The columns are order paramteres that transform similarly under square lattice operations: $\Delta(k)\rightarrow\pm\Delta(k)$ under rotation by $\pi/2$ and reflection across $x,y$ axis. Only the second row can appear in the co-existense phase discussed in the text, in which pairing is between electrons on opposite sub-lattices.}
   \label{fig:orderp}
 \end{figure}

The remainder of this subsection will characterize the symmetry properties of the possible
superconducting phases.
We also allow long-range N\'eel order by a condensation of the CP$^1$ fields with $\langle z^{\alpha*}z_\beta\rangle = \delta_{\alpha \beta} m_\alpha$: the N\'eel order is then polarized in the $z$ direction with
spontaneous moment $m_{\uparrow}-m_{\downarrow}$. Using these averages, Eq.~(\ref{eq:gpair}), and the expressions for the physical electron operators, Eq.~(\ref{eq:physel}), we can calculate the required anomalous averages,
\begin{eqnarray}
\label{eq:orderp}
\langle \Psi_{1\alpha}(k) \Psi_{1\beta}(-k)\rangle &=& -\varepsilon_{\alpha\beta}[m_\beta \Delta(k) + m_\alpha \Delta(-k)] \nonumber\\
\langle \Psi_{1\alpha}(k) \Psi_{2\beta}(-k)\rangle &=& \varepsilon_{\alpha\beta}[m_\beta \Delta(k) - m_\alpha \Delta(-k)] \nonumber\\
\langle \Psi_{2\alpha}(k) \Psi_{2\beta}(-k)\rangle &=& \varepsilon_{\alpha\beta}[m_\beta \Delta(k) + m_\alpha \Delta(-k)]
\end{eqnarray}
At the critical point from the AFM+SC state to the SC state the N\'eel order parameter vanishes, i.e., $m_{\uparrow}=m_{\downarrow}$ and the $\Psi_1\Psi_2$ correlator should disappear (this follows from the restoration of full translational invariance), indicating that $\Delta(k)=\Delta(-k)$. Since the superconducting instability arises out of a short-range attractive interaction it is most natural to expect $s$-wave pairing. Remarkably, this naturally leads to $d_{x^2-y^2}$ pairing for the physical electrons [as can be verified from Eq.~(\ref{eq:orderp}) by substituting $\Delta(k)=\Delta_0$]. However since the underlying $g_\pm$ particles are in an $s$-wave state, the quasi-particles in this $d_{x^2-y^2}$ superconductor are fully gapped. We propose that this is the quantum state that describes the phase AFM+SC in Fig.~\ref{fig:Tx} and that is observed in the region of co-existence in \NCCO~\cite{greven07}. We note that with increasing $x$, the N\'eel order is suppressed making the gauge field mediated attraction (that causes superconductivity) stronger, which in turn is expected to result in an increase of $T_c$, consistent with experimental observations. For the sake of completeness, we present the other symmetry allowed options for pairings (see Fig.~\ref{fig:orderp} second row): $\Delta(k)=k_x^2-k_y^2$ corresponds to the $s$ case, $\Delta(k)= k_xk_y (k_x^2-k_y^2)$ corresponds to the $d_{xy}$ case and  $\Delta(k)=k_xk_y$ corresponds to the $g$ case; all these states have nodal excitations.
Finally, the condition $\Delta(k+K_N)=-\Delta(k)$ that is satisfied by all the order parameters deduced from Eq.~\ref{eq:orderp} (and that are illustrated in the second row of Fig.~\ref{fig:orderp}) follows quite simply from the fact that the phase factor $e^{iK_N \cdot r}$ is $+1(-1)$ on the $A(B)$ sublattices and that pairing occurs only between electrons on opposite sub-lattices.

\section{Quantum criticality}
\label{sec:qc}

We now turn to our main results on the quantum phase transitions involving loss of N\'eel order
as described by the low energy theory introduced in Section~\ref{sec:fields}; the results were summarized
in Fig.~\ref{fig:phases1} and \ref{fig:phases2}.

\subsection{Metallic states}
\label{sec:metal}

First, let us consider the transition without superconductivity, destroying magnetic order in the
AFM metal, leading to the doublon metal. This transition is described directly by the field
theory in Section~\ref{sec:fields}, and is associated with the condensation of the $z_\alpha$ spinons
in the presence of the $g_\pm$ Fermi surfaces.

At $T=0$, such a transition between
metallic states could be induced by destroying superconductivity by an applied magnetic field. Moreover,
even at zero magnetic field, the quantum critical region at temperatures above the superconducting $T_c$
could be controlled by the crossovers of an underlying AFM metal/doublon metal quantum critical point.
Monopoles can be ignored in the following analysis because they are suppressed by the gapless excitations
at the $g_{\pm}$ Fermi surfaces \cite{hermele} (see Appendix~\ref{sec:monopole}). 
The resulting state without antiferromagnetism therefore
carries gapless gauge excitations, and as we noted earlier, realizes an algebraic charge liquid which we call
a doublon metal.

The theory for this transition follows the analysis of a formally similar transition of bosons and fermions coupled
to a U(1) gauge field in Ref.~\onlinecite{ffl}. In this previous case, the bosons were spinless and fermions carried spin,
whereas here the fermions are spinless while the bosons carry spin. However, for the quantum criticality,
the more significant difference is that the quadratic action for the $z_\alpha$ bosons has a relativistic structure, unlike
the $z=2$ dispersion in Ref.~\onlinecite{ffl}.

The renormalized $A_\mu$ gauge field propagator is a key ingredient in our analysis. This depends upon
the polarizabilities of the $g$ fermions and the $z_\alpha$ bosons at the quantum critical point. We evaluate these
from the bare actions $\mathcal{L}_g$ and $\mathcal{L}_z$, and will confirm later that the same
results hold in the fully renormalized critical theory. As usual, the fermion polarizability screens the longitudinal
$A_\mu$ fluctuations, and the only potential singularity arises from the transverse $A_\mu$ propagator, $D$.
In the Coulomb gauge, this has the low momentum and imaginary frequency form \cite{senthilmott}
\begin{equation}
D_{ij} (k, i\omega)  \sim \left( \delta_{ij} - \frac{k_i k_j}{k^2} \right) \frac{1}{k + \chi |\omega|/k}
\label{da}
\end{equation}
Here the $|\omega|/k$ term in the denominator is the contribution of the $g$ fermions, and is present for $|\omega | < v_F k$,
where $v_F$ is the Fermi velocity.
The $k$ term emerges from the critical $z_\alpha$ correlator (it coefficient is proportional to the critical
conductivity of the $z_\alpha$'s).

Let us now compute the consequence of the overdamped gauge fluctuations in Eq.~(\ref{da}) on the $z$ spectral
function. At leading order the $z_\alpha$ self energy at criticality is
\begin{equation}
\Sigma_z (p, i \epsilon)  \sim \int d \omega \int d^2 k \frac{(p^2 - (p\cdot k)^2 /k^2)}{k + \chi |\omega|/k} \frac{1}{(\omega+ \epsilon)^2 + (k+p)^2}. \label{sigmaz}
\end{equation}
It is now easy to confirm that this expression for $\Sigma_z$ is non-singular at low $p$ and $\epsilon$, and does not
modify the leading behavior of the $z$ propagator. In particular, the on-shell self energy has the imaginary part
\begin{equation}
\mbox{Im} \Sigma_z (p, \epsilon = p) \sim p^3,
\end{equation}
which is clearly unimportant to the critical theory. Thus the overdamping of the gauge fluctuations by the $g_{\pm}$ fermions
strongly suppresses their influence on the $z_\alpha$ excitations. Indeed, in the $z=1$ scaling, the $k$ term in the denominator
of Eq.~(\ref{da}) can be neglected, and the renormalized
action for the transverse component of the gauge field is $\sim A_T^2 (|\omega|/k)$; this scales as an anisotropic ``mass'' term for
the gauge boson. Thus we can view this feature as a fermionic version of the Higgs mechanism, in which the low energy
excitations of a Fermi surface quench the gauge field fluctuations. We will comment further on this analogy with the Higgs
mechanism in Section~\ref{sec:bosons}.

In a recent work \cite{senthilmott} in a different context, Senthil has computed the consequences of the singular interactions associated 
with Eqs. (\ref{da}) and (\ref{sigmaz}) on the spectral function of the fermions, and the associated formation of critical
Fermi surfaces \cite{senthilcritical}. All those results apply here too to our theory of the transition from the AF metal to the doublon metal.

At this point, we are now prepared to integrate out the $A_\mu$ gauge boson and the $g_\pm$ fermions, and obtain
an effective theory for the $z_\alpha$ spinons. Keeping only the terms potentially relevant near the critical point,
the resulting effective action has the structure
\begin{equation}
\mathcal{S}_z^{\rm eff} = \int d^2 r d\tau \left[ |\partial_\mu z_\alpha|^2 + s |z_\alpha|^2 + u \left(|z_\alpha|^2 \right)^2 \right]
+ \lambda \int d^2 k d \omega [ |z_\alpha|^2 ]{(-k, -\omega)} \, \frac{|\omega|}{k} \, [ |z_\beta|^2 ]{(k, \omega)}
\label{eq:o4}
\end{equation}
The last $\lambda$ term is a consequence of the compressible fluctuations of the $g_{\pm}$ Fermi surfaces, which couple
to $|z_\alpha|^2$ via a contact term \cite{ffliq}. At $\lambda=0$ it is now evident that $\mathcal{S}_z^{\rm eff} $ describes
a transition for the loss of N\'eel order by a conformal field theory in the O(4) universality class. We can therefore ask for the scaling
dimension of $\lambda$ at this conformal critical point. This follows from a simple scaling argument \cite{morinari}:
\begin{equation}
\mbox{dim}[\lambda] = -3 + \frac{2}{\nu}
\end{equation}
The O(4) model has  \cite{zj} $\nu = 0.733$ and so $\lambda$ is an irrelevant perturbation. Further, when we account
for the long-range Coulomb interactions between the $g_{\pm}$ fermions, there is an additional factor of $k$ in the $\lambda$
term, and $\lambda$ is then more strongly irrelevant.

We have now established that the transition from the N\'eel-ordered Fermi-pocket metal to the doublon metal is
in the O(4) universality class. The N\'eel order parameter itself is a quadratic composite of the $z_\alpha$. It transforms
under the symmetric, traceless, second-rank tensor representation of O(4), and the scaling dimension of this composite
operator has been computed earlier \cite{vicari,kim}. From the field-theoretic analysis of Calabrese {\em et al.} \cite{vicari}
we find $\eta_N = 5 - 2 y_{2,2} = 1.374(12)$, while the Monte Carlo simulations of Isakov {\em et al.} \cite{kim} we obtain
$\eta_N = 1.373(3)$.

\subsection{Superconducting states}
\label{sec:supercond}

Now we discuss the transition out of the superconducting
AFM+SC state with increasing doping. Because the $g_{\pm}$
fermions are fully gapped in the superconductor, they can
initially be ignored in the analysis of the critical theory. The
remaining $z_\alpha$ excitations are described by the CP$^1$
model. So a natural initial guess is that the critical theory for the loss of N\'eel order is the
same as that in the insulator \cite{senthil1}. The presence of
superconductivity here does induce additional gapless density
fluctuations, but these are irrelevant \cite{frey} as long as $\nu
> 2/3$ for neutral systems, and generically unimportant with
long-range Coulomb interactions. Further, the paramagnetic state
so obtained is not a BCS superconductor, because the fermionic
Bogoliubov quasiparticles carry no spin. Rather, as discussed in
Ref.~\onlinecite{acl}, it is a ``doublon superconductor''.
However, once we have moved away from the critical point, there
are no gapless excitations which can serve to suppress monopoles
in the U(1) gauge field. We expect that the condensation of the
monopoles at a large secondary length scale will induce
confinement, leading to a generic instability of the doublon
superconductor. 

We are interested in the nature of the confined state.
For the corresponding transition in the insulator \cite{senthil1},
the confining state was the valence bond solid (VBS) which was induced by the Berry phases on the monopoles.
However, here there is the possibility that the density fluctuations of the superconductor can modify
the influence of the Berry phases. Because the $g_\pm$ fermions are paired in the AFM+SC state and at the critical point, it is plausible
that a $t$-$J$ model in which the $g_\pm$ are bosons (and the $z_\alpha$ remain bosons) should
have essentially the same properties in its charge and density correlations in their respective paired states: 
we are merely replacing the internal
constituents of the Cooper pairs, but this should not modify the nature of the phase and vortex fluctuations
of the superfluid.
We will examine such a 
$t$-$J$ model of bosons in Section~\ref{sec:bosons}: we are able to carry out an analysis of the influence
of monopole condensation in some detail, and find 3 distinct possibilities which were listed earlier
in Section~\ref{sec:intro} and Fig.~\ref{fig:phases2}:\\
({\em a\/}) A conventional O(3) transition, as in SDW, theory, to a $d$-wave superconductor with full square lattice symmetry.
The monopole Berry phases are precisely cancelled by density fluctuations in the superfluid, and so the monopoles
confine the $z_\alpha$ spinons into the vector SDW order parameter. \\
({\em b\/}) A `deconfined' CP$^1$ transition to a valence bond supersolid \cite{rsi,vs}, where
the pattern of the VBS order is the same as that in the insulator \cite{ssnature,rsb} (see Fig.~\ref{fig:ins}). Here the monopole Berry 
phases remain as in Eq.~(\ref{eq:sb}). For rational $x/2 = p/q$ with $q/2$ odd, other patterns of order are possible, as discussed in Section~\ref{sec:gammapp}.
\\
({\em c\/}) A direct transition to a $d$-wave superconductor with square lattice symmetry broken as in the states in
Fig.~\ref{q2}, one of which is 
a {\em nematic superconductor\/}. In this case, the monopole Berry phases are only partially compensated by
the superfluid modes, so that monopoles with even magnetic charge are allowed at the transition. Little is known
about the critical properties of such a `doubled monopole' theory, and it is possible the transition is first order.

The above list exhausts the possible transition out of the AFM+SC state, for the case in which
the AFM+SC state does not have density modulations of its own.
In Section~\ref{sec:bosons}, we consider the further possibility that
the AFM+SC state
already carries density modulations (so that it is also a supersolid). We classify transition out of such states:
monopole condensation modifies the nature of the 
density modulations in the non-magnetic supersolid, and these will be described in Section~\ref{sec:max}. 
An interesting feature of the resulting supersolids is that the set of allowed wavevectors for density modulations
are the same as those of a model of paired particles \cite{bbbss1,annals} of density $1+x$. Thus, once N\'eel order is lost,
the primary role of the monopoles is to account for the `background' density of one particle per site in the Mott insulator,
and to combine this density with the doped particles to yield states which are sensitive to the total density.

This sensitivity to the total density in the supersolid bears some similarity to the constraints
placed by Luttinger's theorem on the volume enclosed by the Fermi surface in Fermi liquid states \cite{ffl}.
Let us explain the connection more explicitly.
We can view the monopole
Berry phases as arising from a filled band of anti-holons in the
insulator, and these are extracted into the confining electronic
states \cite{qi}. To see this, let us recall the origin of the monopole Berry phase
in Eq.~(\ref{eq:sb}). This can be traced to the constraint
$b_\alpha^\dagger b_\alpha + g_+^\dagger g_+ = 1$ applied on every
site of sublattice A (there are parallel considerations on
sublattice B, which we will not write down explicitly), and
implemented by a Lagrange multiplier $\lambda$ in the effective
action with the term
\begin{equation}
i \int d \tau \, \lambda ( b_\alpha^\dagger b_\alpha + g_+^\dagger g_+ - 1).
\label{eq:lambda}
\end{equation}
The fluctuations of $\lambda$ are $A_\tau$ on sublattice A ($-A_\tau$ on sublattice B),
and the $-1$ in the brackets above evaluates \cite{sj} to Eq.~(\ref{eq:sb}) for a monopole configuration of $A_\mu$.
Let us now also allow for the
gapped holon states\cite{ffliq,qi} by the holon operators $f_\pm$, in which case the term in Eq.~(\ref{eq:lambda})
generalizes to
\begin{equation}
i \int d \tau \, \lambda ( b_\alpha^\dagger b_\alpha + g_+^\dagger g_+ + f_+^\dagger f_+ - 1).
\label{eq:lambda2}
\end{equation}
Finally, we perform a particle-hole transformation to anti-holons $h_+ = f_+^\dagger$ (which are distinct
from the doublons) to obtain
\begin{equation}
i \int d \tau \, \lambda ( b_\alpha^\dagger b_\alpha + g_+^\dagger g_+ - h_+^\dagger h_+ ).
\label{eq:lambda3}
\end{equation}
In this form, there is no $-1$ in the bracket, but we have a filled band of $h_+$ anti-holons---thus we have
an alternative
book-keeping in which there is
no monopole Berry phase, but we do have to account for the unit density of anti-holons in the Mott insulator. 
After confinement
with spinons, it is this density which contributes to the expansion to the large Fermi surface in the Fermi liquid,
and the sensitivity of the supersolid to density $1+x$.

Next, we consider the structure of the physical electron spectral function in the supersolid.
We focus on momenta along the diagonals of the square lattice Brillouin zone. Right at the critical point, the $g_\pm$
and $f_\pm$ fermionic excitations are fully gapped, and so the electron spectral function (which is a convolution of these
fermion Green's functions with those of the $z_\alpha$) is also fully gapped. Moving on the confining side of the critical point,
a natural possibility is that electron spectrum remains fully gapped \cite{vojtass,granath,parkss}.
However, given the fact
that the confining supersolid consists of a density $1+x$ (as in a `large' Fermi surface), 
from Fig.~\ref{fig:fs} it would
seem natural that this state has 4 gapless nodal quasiparticles; if so, the total
spectral weight in these low energy fermions would vanish as we approach the critical point, in a manner we expect is related
to the scaling dimension of the monopole operators. It is interesting to note that this vanishing of low energy fermionic
spectral weight resembles the phenomenon of spectral weight transfer in dynamical mean field theory \cite{dsf,dmft}. 
We also note that the emergence of gapless composite fermions from gapful constituents (here the `composite'
electrons consist of gapped spinons and holons) has counterparts in a number of particle theory models \cite{pw,aspy}.

It is useful to discuss this theory in the context of 
recent ideas by Senthil \cite{senthilcritical} on `critical Fermi surfaces'. In the latter framework for a transition to a 
$d$-wave superconductor with 4 nodal points, the
nodal fermions would be part of the critical theory, and then the deconfined critical theory would not
be the CP$^1$ model. Such a scenario would be realized here if the $f_\pm$ holon Fermi surfaces
formed in the AFM state (this is compatible with current photoemission experiments \cite{armitage03}),
and the magnetic disordering transition led to a holon superconductor with gapless Dirac excitations \cite{acl}.
A confinement transition on the holon superconductor would then realize this scenario, but only
at 'critical Fermi points' and not on a 'critical Fermi surface'.

\section{$t$-$J$ model of bosons}
\label{sec:bosons}

The Higgs-like suppression of the $A_\mu$ fluctuations in Section~\ref{sec:metal} suggested to us that
we examine a toy model of bosons obeying the same $t$-$J$ model described here. In other words,
we will consider the same theory presented in Section~\ref{sec:fields} but now the $g_\pm$ and
the $\Psi$ fields are all bosons. We can make quite reliable statements about the phases of this
model, including the role of monopoles and Berry phases.

A further motivation to examining this model was noted in Section~\ref{sec:supercond}: this is an efficient
way to analyze the paired superconducting states, where we expect pairs of bosons or fermions to 
have similar properties.

The analogy between the phases of the electronic model and the toy boson model were summarized in
Fig.~\ref{fig:phases1} and \ref{fig:phases2}. The parallel of the Higgs-like effects in the metallic phases of
Section~\ref{sec:metal} appears when
we replace the $g_\pm$ Fermi pockets by Bose condensates of the $g_\pm$--- the
corresponding transitions in the boson model are then in the same universality class as in the metallic
electronic model. As indicated in Fig.~\ref{fig:phases2}, the boson model also has parallels to the transitions
of the superconducting sector of the electronic model which were discussed in Sections~\ref{sec:supercond}. 
This will be described in more detail below.

First, let us list the phases of the boson $t$-$J$ model of interest to us:\\
({\em i\/}) \underline{AFM boson superfluid}: Here both the $z_\alpha$ and the $g_\pm$ condense with
\begin{equation}
\langle z_\alpha \rangle \neq 0~~,~~\langle g_\pm \rangle \neq 0.
\label{eq:b1}
\end{equation}
The presence of these condensates implies that both $A_\mu$ and monopole fluctuations
are suppressed, as in the AFM metal.
Also, by Eq.~(\ref{eq:physel}), the physical boson operator $\Psi$ also has a non-zero condensate.
So this state has AFM order and a flux quantum of $h/e$.\\
({\em ii\/}) \underline{Paired boson superfluid}: Now spin rotation invariance is restored, with
\begin{equation}
\langle z_\alpha \rangle  = 0~~,~~\langle g_\pm \rangle \neq 0.
\label{eq:b2}
\end{equation}
However, the $g_\pm$ condensate is sufficient to continue to suppress both the $A_\mu$ and the
monopole fluctuations, making this state the analog of the doublon metal. Two other characteristics
of this state reinforce the analogy with the doublon metal: ({\em i\/}) the $z_\alpha$ quanta
represent stable, neutral, $S=1/2$, gapped excitations, which are also found in the doublon metal,
and ({\em ii\/}) the action of an isolated monopole diverges linearly with system size because of the Higgs
condensate, and a similar linear divergence appears \cite{sungsik} in an RPA-like estimation \cite{nagaosa} of the monopole action in the doublon 
metal (see Appendix~\ref{sec:monopole}). 
With the condensates as in Eq.~(\ref{eq:b2}), 
as discussed in Ref.~\onlinecite{annals}, the only gauge-invariant condensate carries charge $2e$,
and so the flux quantum is $h/(2e)$. A comparison of Eqs. (\ref{eq:b1}) and (\ref{eq:b2}) shows that
the transition between the AFM boson superfluid and the paired boson superfluid involves criticality
of $z_\alpha$ alone. The $A_\mu$ mode can be ignored and so it is evident that the critical theory
is the O(4) model in Eq.~(\ref{eq:o4}), but with the last density fluctuation term replaced by
the analogous term for a superfluid \cite{frey}.
The latter term is also irrelevant, by an argument similar to that made for the
electronic case.\\
({\em iii\/}) \underline{AFM paired boson superfluid}: Now we condense the $z_\alpha$, but only allow
for a paired condensate of the $g_\pm$ bosons with
\begin{equation}
\langle z_\alpha \rangle  \neq 0~~,~~\langle g_\pm \rangle = 0~~,~~\langle g_+ g_- \rangle \neq 0.
\label{eq:b3}
\end{equation}
There is antiferromagnetic order, and the flux quantum is $h/(2e)$. The $z_\alpha$ condensate is sufficient
to suppress both the $A_\mu$ and the
monopole fluctuations, making this state the analog of the AFM superconductor in the electronic model.
In some cases (to be discussed below in Section~\ref{sec:max}) this state will also break translational symmetry,
i.e. it will become a supersolid.
\\
({\em iv\/}) \underline{Paired boson supersolid}: The only condensate is that associated with the paired bosons:
\begin{equation}
\langle z_\alpha \rangle  = 0~~,~~\langle g_\pm \rangle = 0~~,~~\langle g_+ g_- \rangle \neq 0.
\label{eq:b4}
\end{equation}
This is the most interesting state here: the $A_\mu$ and monopole fluctuations are not suppressed,
and we expect a crossover to a confining state. The same phenomenon also appeared in the electronic
case with the doublon superconductor, which we argued was unstable to confinement to a $d$-wave
superconductor. The key advantage of the toy boson model is that we can describe the crossover
to confinement in some detail, as will be presented in the following subsections. Our main result will
be that there are generally periodic bond/density modulations in this phase, {\em i.e.\/} it is a supersolid;
we include here the case of the nematic superconductor, in which only the $Z_4$ rotational symmetry
of the square lattice is broken.
Finally, we will demonstrate that these
modulations are characteristic \cite{bbbss1,annals} of the total density of bosons, $1+x$.

\subsection{Duality and symmetry analysis}
\label{sec:dual}

We will apply the analog of the duality methods presented in Refs.~\onlinecite{senthil1,bbbss1,annals}
to this model. These dualities are only operative for abelian symmetry, and so we shall replace the
SU(2) spin symmetry by a U(1) symmetry of spin rotations about the $z$ axis.

We write the spinons, $z_\alpha$, and represent them by two angular degrees
of freedom $z_\uparrow = e^{i \theta_\uparrow}$, $z_\downarrow = e^{i \theta_\downarrow}$.
Similarly we take the $g_{\pm}$ (which are now bosons) and write them as
$g_{\pm} = e^{i \phi_{\pm}}$. These fields are coupled to a compact U(1) gauge
field $A_\mu$, with the same charges as in the body of the paper. Finally, the monopoles
in $A_\mu$ are endowed with the Haldane Berry phases \cite{sj,senthil1} in Eq.~(\ref{eq:sb}), to
properly include the physics of the insulating antiferromagnet.

The simplest model consistent with such a framework is written below.
Here we have discretized spacetime onto the sites of direct cubic lattice with sites $j$ 
and $\Delta_\mu$ is a discrete lattice derivative.
\begin{eqnarray}
\mathcal{Z} &&=  \prod_j \int d\theta_{\uparrow j} d\theta_{\downarrow j} d \phi_{+j} d \phi_{- j}  dA_{j \mu}  \exp \Biggl( \nonumber \\
&&~~   \frac{1}{K} \sum_{j \mu} \cos \left( \Delta_\mu \theta_{\uparrow j} - {A}_{j\mu}   \right) + \frac{1}{K} \sum_{j \mu} \cos \left( \Delta_\mu \theta_{\downarrow j}  - {A}_{j\mu}    \right)  \nonumber \\
&&~~ + \frac{1}{L} \sum_{j \mu} \cos \left( \Delta_\mu \phi_{+ j} - {A}_{j\mu}  - B_{j\mu} \right) + \frac{1}{L} \sum_{j \mu} \cos \left( \Delta_\mu \phi_{- j}  + {A}_{j\mu} -B_{j \mu}   \right) \nonumber \\
&&~~ + \frac{1}{e^2}
\sum_{\Box} \cos \left( \epsilon_{\mu\nu\lambda}\Delta_{\nu} {A}_{j
\lambda} 
\right) - \mathcal{S}_B  \Biggr).
\label{zzz0}
\end{eqnarray}
Apart from the coupling constants, $K$, $L$, $e^2$, the action contains two fixed external fields.
The uniform static external electromagnetic  field ${B}_\mu = i \overline{\mu} \delta_{\mu\tau}$,
where $\overline{\mu}$ is the chemical potential; the value of $\overline{\mu}$ is adjusted so that
density of each $g_\pm$ boson species is $x/2$. The last term accounts for the Berry phases linked
to the monopoles in $A_\mu$ by Eq.~(\ref{eq:sb}).

To be complete, we should also add to Eq.~(\ref{zzz0}) a staggered chemical potential which preferentially
locates the $g_\pm$ on opposite sublattices, as has been done in previous work \cite{bbbss2,annals}. However,
this term is not essential for our conclusions here, and so we omit it in the interests of simplicity.

The duality analysis of Eq.~(\ref{zzz0}) is most transparent when the action is written in a Villain (periodic Gaussian) form.
We do this by introducing the integer-valued fields, $p_{\uparrow j\mu}$, $p_{\downarrow j\mu}$, $n_{+j\mu}$,
$n_{-j\mu}$ which reside on the links of the direct lattice, and the integer-valued $q_{\jmath\mu}$ which resides on the links
of the dual lattice. The dual lattice sites are labeled by $\jmath$.
\begin{eqnarray}
\mathcal{Z} &&= \sum_{\{p_{\uparrow j\mu}\}} \sum_{\{p_{\downarrow j\mu}\}}
\sum_{\{n_{+ j\mu}\}} \sum_{\{n_{- j\mu}\}} \sum_{ \{q_{\jmath \mu} \}}  \prod_j \int d\theta_{\uparrow j} d\theta_{\downarrow j} d \phi_{+j} d \phi_{- j}  dA_{j \mu}  \exp \Biggl( \nonumber \\
&&~~  - \frac{1}{2K} \sum_{j \mu} \left( \Delta_\mu \theta_{\uparrow j} - {A}_{j\mu}  - 2 \pi p_{\uparrow j\mu} \right)^2 - \frac{1}{2K} \sum_{j \mu} \left( \Delta_\mu \theta_{\downarrow j}  - {A}_{j\mu}   - 2 \pi p_{\downarrow j\mu} \right)^2  \nonumber \\
&&~~ - \frac{1}{2L} \sum_{j \mu} \left( \Delta_\mu \phi_{+ j} - {A}_{j\mu}  - B_{j\mu} - 2 \pi n_{+ j\mu} \right)^2 - \frac{1}{2L} \sum_{j \mu} \left( \Delta_\mu \phi_{- j}  + {A}_{j\mu} -B_{j \mu}  - 2 \pi n_{- j\mu} \right)^2 \nonumber \\
&&~~ - \frac{1}{2e^2}
\sum_{\Box} \left( \epsilon_{\mu\nu\lambda}\Delta_{\nu} {A}_{j
\lambda} - 2 \pi q_{\jmath \mu}
\right)^2  - \frac{i \pi}{2} \sum_\jmath \zeta_\jmath \Delta_{\mu} q_{\jmath \mu}\Biggr).
\label{zzz}
\end{eqnarray}
An advantage of this periodic Gaussian form is that we are able to write an explicit expression for the monopole Berry phase \cite{sj};
the fixed field $\zeta_\jmath = 0,1,2,3$ is the same as
that appearing in Eq.~(\ref{eq:sb}).

Now we proceed with a standard duality transformation of this action. Initially, this maps the theory
onto  the integer valued spin currents $J_{\uparrow j \mu}$ and $J_{\downarrow j\mu}$, the integer valued
charge currents $H_{+j\mu}$ and $H_{-j \mu}$,  and the integer valued fluxes $Q_{\jmath \mu}$
with the partition function
\begin{eqnarray}
\mathcal{Z}_d &&= \sum_{\{J_{\uparrow j\mu}\}} \sum_{\{J_{\downarrow j\mu}\}}
\sum_{\{H_{- j\mu}\}} \sum_{\{H_{- j\mu}\}} \sum_{ \{Q_{\jmath \mu} \}} \delta_{\rm constraints}
\exp \Biggl(
- \frac{K}{2} \sum_{j\mu} \left( J_{\uparrow j \mu}^2 + J_{\downarrow j \mu}^2 \right)\nonumber \\
&& - \frac{L}{2} \sum_{j\mu} \left( H_{+j \mu}^2 + H_{-j \mu}^2 \right)  - \frac{e^2}{2} \sum_{\jmath \mu} \left( Q_{\jmath \mu} - \frac{1}{4} \Delta_\mu \zeta_{\jmath \mu} \right)^2 - i \sum_{j\mu}
{B}_{j\mu} \left( H_{+j\mu} + H_{- j \mu} \right) \Biggr).
\end{eqnarray}
The summations in $\mathcal{Z}_d$ are restricted to integer-valued fields which obey the local constraints
\begin{eqnarray}
&& \Delta_\mu J_{\uparrow j\mu} = 0~~,~~\Delta_\mu J_{\downarrow j\mu} = 0~~,~~
\Delta_\mu H_{+\mu} = 0~~,~~\Delta_\mu H_{-j\mu} = 0~~,\nonumber \\
&&~~~~~~~~~\epsilon_{\mu\nu\lambda} \Delta_\nu
Q_{\jmath \lambda} = J_{\uparrow j\mu} + J_{\downarrow j\mu} + H_{+j\mu} - H_{-j\mu}.
\end{eqnarray}
We solve these constraints by introducing the dual gauge fields $a_{\uparrow \jmath\mu}$ and $a_{\downarrow \jmath\mu}$ whose fluxes are the spin currents, the dual gauge fields $b_{+ \jmath\mu}$ and  $b_{- \jmath \mu}$
whose fluxes are the charge currents, and a height field $h_{\jmath}$ whose gradients are the $A_\mu$ fluxes. Finally, we promote these
dual discrete fields to continuous fields by introducing the dual matter fields $e^{- i \alpha_{\uparrow\jmath}}$
and $e^{- i \alpha_{\downarrow \jmath}}$ which annihilate vortices in the $z_{\uparrow,\downarrow}$ spinons, the dual
matter fields $e^{- i \beta_{+\jmath}}$
and $e^{- i \beta_{- \jmath}}$ which annihilate vortices in the $g_{\pm}$ charged bosons,
and the corresponding vortex and monopole fugacities. This leads to the dual
theory in its unconstrained form
\begin{eqnarray}
\mathcal{Z}_{d2} && = \prod_j \int d a_{\uparrow \jmath \mu} d a_{\downarrow \jmath \mu}
d b_{+ \jmath \mu} d b_{- \jmath \mu} dh_{\jmath} d \alpha_{\uparrow \jmath} d \alpha_{\downarrow \jmath} d \beta_{+ \jmath} d \beta_{-\jmath}
 \exp \Biggl(  \nonumber \\
&&~~~~~~~~~~~~  - \frac{K}{2} \sum_{\Box} \left(
\left( \epsilon_{\mu\nu\lambda} \Delta_\nu a_{\uparrow \jmath \lambda} \right)^2   + \left( \epsilon_{\mu\nu\lambda} \Delta_\nu a_{\downarrow \jmath \lambda}\right)^2 \right)
\nonumber \\
&&~~~~~~~~~~~~  - \frac{L}{2} \sum_{\Box} \left(
\left( \epsilon_{\mu\nu\lambda} \Delta_\nu b_{+ \jmath \lambda} - \frac{\overline{\mu}}{L} \delta_{\mu\tau} \right)^2   + \left( \epsilon_{\mu\nu\lambda} \Delta_\nu b_{-\jmath \lambda}- \frac{\overline{\mu}}{L}  \delta_{\mu\tau} \right)^2 \right)
\nonumber \\
&&~~~~~~~~~~~~ - \frac{e^2}{2} \sum_{\jmath \mu} \left(  \Delta_\mu h_{\jmath }
+ a_{\uparrow \jmath \mu} + a_{ \downarrow \jmath \mu}  + b_{+ \jmath \mu} - b_{- \jmath \mu} \right)^2 \nonumber \\
&&~~~~~~~~~~~~ + y_{vs} \sum_{\jmath \mu} \left( \cos \left(\Delta_{\mu} \alpha_{\uparrow\jmath} - 2 \pi a_{\uparrow\jmath \mu} \right)
+ \cos \left(\Delta_{\mu} \alpha_{\downarrow \jmath} - 2 \pi a_{\downarrow \jmath \mu} \right)  \right) \nonumber \\
&&~~~~~~~~~~~~ + y_{vc} \sum_{\jmath \mu} \left( \cos \left(\Delta_{\mu} \beta_{+\jmath} - 2 \pi b_{+\jmath \mu} \right)
+ \cos \left(\Delta_{\mu} \beta_{- \jmath} - 2 \pi b_{- \jmath \mu} \right)  \right) \nonumber \\
&&~~~~~~~~~~~~ + y_m
\sum_{\jmath} \cos \left(2 \pi  h_\jmath + \alpha_{\uparrow\jmath}+ \alpha_{\downarrow\jmath} + \beta_{+\jmath}-\beta_{-\jmath} + \frac{\pi}{2} \zeta_{\jmath} \right) \Biggr). \label{zzzd2}
\end{eqnarray}
The average flux of $b_{\pm}$ is $\overline{\mu}/{L}$ and this should equal half the electron density, $x/2$.

The action in Eq.~(\ref{zzzd2}) appears to be of daunting complexity, but its physical interpretation is
transparently related to the direct theory. There are 4 vortex matter fields, $e^{i \alpha_{\uparrow}}$,
$e^{i \alpha_{\downarrow}}$, $e^{i \beta_+}$, $e^{i \beta_-}$. These annihilate vortices
in $z_\uparrow$, $z_\downarrow$, $g_+$ and $g_-$ respectively. These 4 matter fields
carry unit charges under 4 U(1) gauge fields, $a_\uparrow$, $a_\downarrow$, $b_+$, and $b_-$ respectively.
Of these 4 gauge fields, one combination is always Higgsed out by the scalar field $h$ (by `Higgsed' we mean that the gauge
boson acquires a mass via the Higgs mechanism). The latter is related
to the monopole annihilation operator $e^{2 \pi i h}$, and the monopoles carry Berry phases
$e^{i \pi \zeta_\jmath /2}$.

Let us now make a further simplification of the dual action in Eq.~(\ref{zzzd2}). 
As the gauge field combination $a_{\uparrow} + a_{\downarrow} + b_+ - b_-$ is
always Higgsed by the $h$ field it is convenient to integrate these two fields out, obtaining the dual action
\bea S_d &=& K \sum_{\Box} (\epsilon_{\mu \nu \lambda} \Delta_\nu a_{\lambda})^2 + L \sum_{\Box} (\epsilon_{\mu \nu \lambda} \Delta_\nu b_{\lambda} - \frac{\bar{\mu}}{L} \delta_{\mu \tau})^2 + (K+L) \sum_{\Box}(\epsilon_{\mu \nu \lambda} \Delta_\nu c_\lambda)^2\nn\\
&-&y_{vs} \sum_{\jmath \mu} \left(\cos(\Delta_\mu \alpha_{\uparrow \jmath} - 2\pi a_{\jmath\mu} - 2\pi c_{\jmath\mu})+\cos(\Delta_\mu \alpha_{\downarrow \jmath} + 2\pi a_{\jmath\mu} - 2\pi c_{\jmath\mu})\right)\nn\\&-& y_{vc} \sum_{\jmath \mu} \left(\cos(\Delta_\mu \beta_{+ \jmath} - 2\pi b_{\jmath\mu} + 2\pi c_{\jmath\mu})+\cos(\Delta_\mu \beta_{- \jmath} -2\pi b_{\jmath\mu} - 2\pi c_{\jmath\mu})\right)\nn\\ 
&-& y_m \sum_{\jmath} \cos(\alpha_{\uparrow\jmath}+\alpha_{\downarrow\jmath} + \beta_{+\jmath} - \beta_{-\jmath} - \frac{\pi}{2} \zeta_{\jmath})\label{Sduallat}\eea
The resulting action has three gauge fields: $a, b$ and $c$. The flux of $a$ is related to the magnetization density, $S^{z} = \epsilon_{\tau \nu \lambda} \Delta_{\nu} a_{\lambda}$, the flux of $b$ to the electron pair density, $n/2 = \epsilon_{\tau \nu \lambda} \Delta_{\nu} b_{\lambda}$. Finally the field $c$ introduces interactions between spinon and doublon vortices. When $c$ is Higgsed out (as happens in the paired boson superfluids),
the $e^{i \beta_\pm}$ are the physical vortices in the superconducting order parameter \cite{annals} 
which carry flux $h/(2e)$; otherwise they carry flux $h/e$.
As usual, gauge invariant local operators in the direct picture correspond to monopole operators of the dual gauge fields (see  Table \ref{MonTable}).
\begin{table}[t]
\begin{tabular}{|c|c|c|c|}
\hline
\quad \quad &\,\, $q_a$\,\, & \,\,\,$q_b$\,\,\, & \,\,$q_c$ \,\,\\
\hline
\quad $z^{\dagger}_{\uparrow} z_{\downarrow}$\quad\quad & $1$ & $0$ & $0$\\
\hline
$g^{\dagger}_+ g^{\dagger}_-$ & $0$ & $1$ & $0$\\
\hline
$z^{\dagger}_s  g^{\dagger}_-$ & $\frac{1}{2} s$ & $\frac{1}{2}$ & $\frac{1}{2}$\\
\hline
$z_s g^{\dagger}_+$ & $ - \frac{1}{2} s$ & $\frac{1}{2}$ & $-\frac{1}{2}$\\
\hline
\end{tabular}
\caption{Correspondence between local operators in direct theory and monopole operators in dual theory. $q_a$, $q_b$ and $q_c$ are monopole fluxes assoicated with gauge fields $a$, $b$ and $c$ respectively. The subscript $s$ labels spin ($s = 1$ for $\uparrow$ and $s = -1$ for $\downarrow$)}.
\label{MonTable}
\end{table}

For notational convenience below, we define the spinon vortices $\psi_{\uparrow,\downarrow}$ by
\begin{equation}
\psi_\uparrow = e^{i \alpha_\uparrow}~~~~,~~~~\psi_\downarrow = e^{i \alpha_\downarrow}.
\end{equation}

\subsubsection{Symmetries}
\label{sec:symmetries}

A crucial part of our analysis will be an understanding of the symmetries of the action in Eq.~(\ref{Sduallat}). 

First let us consider the action of the space group symmetry of the square lattice. Following the analyses of Refs.~\onlinecite{bbbss1}, \onlinecite{annals}, we will consider the operations $T_{x,y}$ (translation by one lattice site in the $x,y$ directions), $R_{\pi/2}^{\rm dual}$ (rotation by a 90$^\circ$ about a dual lattice site), $I_x^{\rm dual}$ (reflection
$x \rightarrow -x$, with the origin on a dual lattice site), and $\mathcal{T}$ (time-reversal). The action in Eq.~(\ref{Sduallat})
is invariant under these operations with the transformations:
\bea T_x: \,\, && \psi_{\uparrow} \to i \psi^{\dagger}_{\downarrow},\quad\quad \psi_{\downarrow} \to i \psiud\nn\\
            &&e^{i \beta_+} \to e^{i \beta_-},\quad e^{i \beta_-} \to e^{i \beta_+}\nn\\
     T_y: \,\, && \psiu \to \psidd,  \quad\quad \psid \to \psiud\nn\\
               && e^{i \beta_+} \to e^{i \beta_-}, \quad e^{i \beta_-} \to e^{i \beta_+}\nn\\
     R^{\mathrm{dual}}_{\pi/2}:\,\, && \psiu \to e^{i \pi/4} \psidd, \quad  \psid \to e^{i \pi/4} \psiud\nn\\
     						&& e^{i \beta_+} \to e^{i \beta_-}, \quad e^{i \beta_-} \to e^{i \beta_+}\nn\\
     I^{\mathrm{dual}}_x: \,\,&& \psiu \to \psid, \quad\quad \psid \to \psiu \nn\\
     											&& e^{i \beta_+} \to e^{-i \beta_-}, \quad e^{i \beta_-} \to e^{-i \beta_+}\nn\\
     {\cal T}: \,\, && \psiu \to \psid,\quad\quad \psid\to \psiu\nn\\
     							&& 	e^{i \beta_+} \to e^{i \beta_+}, \quad e^{i \beta_-} \to e^{i \beta_-}\label{PSGpsi}
               \eea
The non-trivial transformations of the spinon vortices above are a consequences of the monopole Berry phases, $\zeta_\jmath$
in Eq.~(\ref{Sduallat}).
               
We will be interested below in taking the continuum limit of the effective action for these fields.
Here we have to be careful about the fate of the Cooper pair/doublon vortex fields $e^{i \beta_{\pm}}$.               
Indeed, our vortex fields $e^{i \beta_\pm}$ are propagating in the background of an average flux for the $b$ field that is dual to a finite electron density $x/2$. We will work at a rational density,
\beq \frac{x}{2} = \frac{p}{q}\label{dopingpq}\eeq
where $p$ and $q$ are relatively prime integers, and then (as discussed at length in Ref. \onlinecite{bbbss1}) there are $q$ degenerate minima in the Hofstadter dispersion. We label the vortex excitations at these minima by the complex fields $\varphi_{\pm l}$, with $l = 0,1,2,...q-1$. Thus, in the continuum limit, the vortex fields $e^{i \beta_\pm}$ are replaced by the $2 q$ fields $ \varphi_{\pm l}$.     
Moreover, once the fields $e^{i\beta_{\pm}}$ split into $\varphi_{\pm l}$ multiplets, the transformations become even more non-trivial due to the presence of a background flux of the $b$ field,
\bea T_x: \,\, && \varphi_{a l} \to \varphi_{\bar{a}, l+1}\nn\\
     T_y: \,\, && \varphi_{a l} \to \omega^{-l} \varphi_{\bar{a}l}\nn\\
     R^{\mathrm{dual}}_{\pi/2}:\,\, && \varphi_{al} \to \frac{1}{\sqrt{q}}\sum_{m=0}^{q-1} \varphi_{\bar{a}m} \omega^{l m}\nn\\
     I^{\mathrm{dual}}_x: \,\,&& \varphi_{a l} \to \varphi^{\dagger}_{\bar{a},-l} \nn\\
     {\cal T}: \,\, && \varphi_{a l} \to \varphi_{a l} \label{PSGphi}
     \eea
where all indices are implicitly determined modulo $q$, the ``bar"  operation exchanges $+ \leftrightarrow -$ and,
\beq \omega \equiv e^{2 \pi i p/q}\eeq
Notice that for the transformations in Eq. (\ref{PSGphi}) we have,
\beq T_y T_x = \omega T_x T_y\eeq
and this algebra is crucial\cite{bbbss1} in ensuring the $q$-fold degeneracy of the vortex states. The factor $\omega$ is understood as a transformation in the $U(1)_b$ symmetry group (see below). 

In addition to the space group operations, we should also consider the symmetries associated
with global parts of the 3 U(1) gauge groups. 
More explicitly, we define the transformations in the global parts of the gauge groups as,
\bea (e^{i \theta})_a :\,\, &&\psiu \to e^{i \theta} \psiu,\quad \psid \to e^{-i \theta} \psid\label{U1a}\\
 (e^{i \theta})_b: \,\, &&\varphi_{+l} \to e^{i \theta} \varphi_{+ l},\quad \varphi_{-l} \to e^{i \theta} \varphi_{- l}\\
 (e^{i \theta})_c: \,\, && \psiu \to e^{i \theta} \psiu,\quad \psid \to e^{i \theta} \psid\nn\\
 												&& \varphi_{+ l} \to e^{-i \theta} \varphi_{+ l}, \quad \varphi_{- l} \to e^{i \theta} \varphi_{- l}
 \eea
Then, combining the transformations of spinon vortices Eq. (\ref{PSGpsi}) and Cooper pair/doublon vortices Eq. (\ref{PSGphi}),
\beq T_y T_x = (-1)_c (-\omega)_b T_x T_y\eeq
with all the other relations in the lattice group as in the non-projective case.

We have now enumerated all the symmetries which will determine the structure of the effective
action and the phases. However, these symmetries are still somewhat cumbersome, and it is useful
now to define certain bilinears whose transformation properties are somewhat simpler. 

First, we define bilinears of the Cooper pair/doublon vortices. 
We introduce a set of pair vortex operators\cite{bbbss1},
\beq \gamma_{m n} = \omega^{m n/2} \sum_{l=0}^{q-1} \varphi^{\dagger}_{-l} \varphi_{+, l+n}\, \omega^{l m}
\label{defgamma}
\eeq
with the transformation properties,
\bea T_x :\,\, &&\gamma_{m n}\to \omega^{-m} \gamma^{\dagger}_{-m, -n}\nn\\
T_y :\,\, &&\gamma_{mn} \to \omega^{-n} \gamma^{\dagger}_{-m, -n}\nn\\
R^{\mathrm{dual}}_{\pi/2}: \,\, &&\gamma_{mn} \to \gamma^{\dagger}_{-n, m}\nn\\
I^{\mathrm{dual}}_x : \,\, &&\gamma_{m n} \to \gamma_{-m,n}\nn\\
{\cal T} :\,\,&& \gamma_{m n} \to \gamma_{m n} \nn \\
\nn\\
(e^{i \theta})_a :\,\, && \gamma_{m n} \to \gamma_{m n} \nn \\
(e^{i \theta})_b :\,\, && \gamma_{m n} \to \gamma_{m n} \nn \\
(e^{i \theta})_c :\,\, &&  \gamma_{m n} \to e^{-2i \theta} \gamma_{m n} \label{PSGrho}
\eea
Note that the space group transformations of the $\gamma_{mn}$ are just those of the physical particle density
operator at the wavevector $(2\pi p/q) (m,n)$. However, a crucial point is that $\gamma_{mn}$ is {\em not\/} equivalent
to the particle density operator: this is a consequence of the non-trivial transformation of $\gamma_{mn}$ under
U(1)$_c$ above. Only combinations which are neutral under $U(1)_c$ are physically observable.

Next, we consider the following bilinear of the spinon vortices
\beq \psi = \psi_\uparrow \psi_\downarrow \eeq
which has the transformations
\bea T_x:\,\,&& \psi \to - \psi^{\dagger}\nn\\
		 T_y:\,\, && \psi \to \psi^{\dagger}\nn\\
		 R^{\mathrm{dual}}_{\pi/2}:\,\, && \psi \to i \psi^{\dagger}\nn\\
		 I^{\mathrm{dual}}_x:\,\, && \psi \to \psi\nn\\
		 {\cal T}: \,\, && \psi \to \psi\nn\\
		 R^{\mathrm{dir}}_{\pi/2}:\,\, && \psi \to i \psi \nn \\
		 \nn\\
(e^{i \theta})_a :\,\, && \psi \to \psi \nn \\
(e^{i \theta})_b :\,\, && \psi \to \psi \nn \\
(e^{i \theta})_c :\,\, && \psi \to  e^{2i \theta} \psi 
				 \label{PSGmon}\eea
We have also listed above the transformation for direct lattice rotations,
which follows from the other results.
Notice again that the transformation properties of $\psi$ under the space group
are identical to the VBS observable. However, because of the non-zero charge of $\psi$
under U(1)$_c$, we {\em cannot\/} generically identify $\psi$ with the VBS order.
		 
Finally, we note that the product of $\gamma_{mn}$ and $\psi$
\begin{equation}
\rho_{mn} = \gamma_{mn} \psi , \label{defrho}
\end{equation}		 
is indeed invariant under all the global U(1)'s, and so is the simplest composite operator which
can serve as a density operator. From the space group transformations in Eqs.~(\ref{PSGrho}) and (\ref{PSGmon}),
we observed that $\rho_{mn}$ transforms as a linear combination of the components of the density at
wavevectors $(2\pi p/q)(m,n) + (\pi, 0)$ and $(2\pi p/q)(m,n) + (0,\pi)$.

The relationship in Eq.~(\ref{defrho}) is central to all our results: only the product of the monopole operator $\psi$, and the vortex-anti-vortex
composites represented by the $\gamma_{mn}$, is a physical observable. The requirement that we must consider the product of these
dual operators can be traced to the constraint in Eq.~(\ref{eq:lambda}) in the direct theory. There we noted that the monopole Berry 
phase was tied to the constraint on the sum of the spinon and doublon densities. In the dual theory $\psi=\psi_\uparrow \psi_\downarrow$
accounts from the spinon contribution, while $\gamma_{mn}$ accounts for the density fluctuations in the paired doublon superfluid.
		 
\subsubsection{Continuum theory}
\label{sec:continuum}

We are now faced with the relatively straightforward task of writing down the most general action for the
Cooper pair/doublon vortices, $\varphi_{a\ell}$ and the spinon vortices $\psi_{\uparrow,\downarrow}$, consistent with all
the symmetries enumerated in Section~\ref{sec:symmetries}.

The quadratic kinetic terms are a direct transcription of the terms in Eq.~(\ref{Sduallat}), and lead to the Lagrangian
\bea
L_0 &=& |(\partial_\mu - 2 \pi i a_\mu - 2 \pi i c_\mu ) \psi_\uparrow |^2 + 
|(\partial_\mu + 2 \pi i a_\mu - 2 \pi i c_\mu ) \psi_\downarrow |^2  \nn \\
&+& |(\partial_\mu - 2 \pi i b_\mu + 2 \pi i c_\mu ) \varphi_{+\ell} |^2 + 
|(\partial_\mu - 2 \pi i b_\mu - 2 \pi i c_\mu ) \varphi_{-\ell} |^2  + \ldots
\eea

Most crucial for our purposes will be the terms which directly couple the spinon vortices with the $\varphi_{a\ell}$
vortices. These are most directly deduced from Eqs.~(\ref{PSGrho}) and (\ref{PSGmon}). Clearly, we need a combination
of the $\gamma_{mn}$ which transforms like the VBS operator under the space group operations so that the product with $\psi$
will be invariant under U(1)$_c$ and also under the space group.
Such terms can only be constructed for even $q$, and were considered in Ref.~\onlinecite{annals}; in our present notation 
the simplest term is
\beq
L_1 = \lambda_1 \Bigl[ \psi \left( \gamma_{\pi0} - i \gamma_{0\pi} \right) + \mbox{H.c.} \Bigr] \label{eq:l1}
\eeq
Here we have labelled $\gamma_{mn}$ by a subscript which identifies 
the associated wavevector $(2 \pi p/q) (m,n)$, and will frequently
use this notation below. A higher order term which will be important later is
\beq 
L_2  = \lambda_2 \Bigl[ \psi^2 \left( \gamma_{\pi0}^2 -  \gamma_{0\pi}^2 \right) + \mbox{H.c.} \Bigr] \label{eq:l2}
\eeq

\subsubsection{Phases}
\label{sec:phases}

We are now ready to use the vortex degrees of freedom to
identify and characterize the phases introduced at the beginning of
Section~\ref{sec:bosons}:\\
({\em i\/}) \underline{AFM boson superfluid}: Both the $z_\alpha$ and the $g_\pm$ are condensed,
and so all the vortex fields are gapped:
\begin{equation}
\langle \psi_\uparrow \rangle = 0~~,~~\langle \psi_\downarrow \rangle = 0~~,~~\langle \varphi_{+\ell} \rangle = 0~~,~~\langle \varphi_{- \ell} \rangle = 0.
\label{eq:b5}
\end{equation}
We will also need to consider independent condensates of bilinears of the vortices in the charges $g_\pm$ below,
and so let us also note that in this phase
\begin{equation}
\langle  \varphi_{- m}^{\dagger} \varphi_{+\ell} \rangle = 0~~,~~
\langle \varphi_{- m} \varphi_{+\ell} \rangle = 0. \label{eq:phiphi}
\end{equation}
The low energy excitations
of this phase consist of the 3 U(1) photons, $a$, $b$, $c$. These 3 photons
correspond to the 3 spin-wave modes that are easily deduced to be present in this phase of the direct theory.\\
({\em ii\/}) \underline{Paired boson superfluid}: The restoration of the spin rotation invariance implies
that the vortices in the spinons $z_\alpha$ have condensed:
\beq
\langle \psi_\uparrow \rangle \neq 0~~,~~\langle \psi_\downarrow \rangle \neq 0~~,~~\langle \varphi_{+\ell} \rangle = 0~~,~~\langle \varphi_{- \ell} \rangle = 0.
\eeq
The condensation of $\psi_{\uparrow \downarrow}$ implies that we also have $\langle \psi \rangle \neq 0$.
However, this does not imply the appearance of VBS order, or broken translational symmetry, because
of the non-zero U(1)$_c$ charge carried by $\psi$. Note also that because of the coupling in Eq.~(\ref{eq:l1}),
that a particular bilinear of the $\varphi_{\pm \ell}$ vortices has a non-zero condensate
\beq
 \left\langle \left( \gamma_{\pi0} - i \gamma_{0\pi} \right) \right \rangle \neq 0. \label{eq:rho0p}
\eeq
Again, this condensate does not break translational symmetry because it has to be combined with $\psi$ to
obtain an observable neutral under U(1)$_c$, and the combination is translationally invariant; indeed this translational
invariance was used to derive the term in Eq.~(\ref{eq:l1}). All other linear combinations of bilinears of the 
$\varphi_{\pm \ell}$ vortices of the form in Eq.~(\ref{eq:phiphi}) have a vanishing expectation value.

We can also use this vortex formulation to analyze the transition between the AFM boson superfluid
and the paired boson superfluid.
Because the vortices $\varphi_{\pm \ell}$ are gapped in both phases, we can set $\varphi_{\pm \ell} = 0$
in all terms in the action. The critical theory then consists of 2 complex scalars $\psi_{\uparrow,\downarrow}$ coupled to 2 U(1) gauge fields $a_{\uparrow} = a + c$ and $a_\downarrow = -a + c$.
The $b$
gauge field is not Higgsed in either phase, and so remains gapless across the transition; this is just
the Goldstone mode of the superfluid order, which is not connected with the critical theory.
We can `undualize' each complex scalar + U(1) gauge field combination: by the Dasgupta-Halperin
duality \cite{dh} this yields a critical theory of 2 complex scalars (and no gauge fields) with O(2)$\times$O(2) symmetry. This critical theory is simply the easy-plane limit of the O(4) theory discussed in the direct formulation above and in Fig.~\ref{fig:phases1}:
it is obtained from the models there by adding the easy-plane anisotropy term $|z_\uparrow|^2
|z_\downarrow |^2$.\\
({\em iii\/}) \underline{AFM paired boson superfluid}:
Like the AFM boson superfluid above, all vortices have a vanishing condensate:
\beq
\langle \psi_\uparrow \rangle = 0~~,~~\langle \psi_\downarrow \rangle = 0~~,~~\langle \varphi_{+\ell} \rangle = 0~~,~~\langle \varphi_{- \ell} \rangle = 0.
\eeq
However, unlike the AFM boson superfluid, we should
only allow for a condensate of the product $g_+ g_-$, and not for the individual boson factors. In the dual variables,
this means at least some linear combinations of the vortex bilinears $\varphi_{- m}^{\dagger} \varphi_{+\ell}$ should have
a non-zero condensate while all the bilinears $\varphi_{- m} \varphi_{+\ell}$ have a vanishing expectation value. 
The simplest choice is to allow
\begin{equation}
\left\langle \gamma_{00} \right\rangle \neq 0 \label{eq:rho00}
\end{equation}
This does not break translational symmetry, and serves the important purpose of Higgsing out the $c$ gauge field
and ensuring that there is no single boson condensate. However it is also possible to choose 
other $\gamma_{mn}$ to have non-zero expectation values. This option will be discussed further in Section~\ref{sec:max},
where we will show that in general such choices do break translational symmetry, and so lead to supersolid 
order (along with the AFM order already present here).
However, there will be a number of other choices, distinct from Eq.~(\ref{eq:rho00}), which do not break translational symmetry.
All these choices lead to AFM boson superfluids which are identical in the sense of symmetry, but do have distinct
`topological order' associated with the alignment of the gauge-dependent condensates of $\varphi_{- m}^{\dagger} \varphi_{+\ell}$.
We will provide a complete listing of all such inequivalent AFM paired boson superfluids in Section~\ref{sec:max}; they
are distinguished, in particular, by distinct universality classes of transitions involving the loss of AFM order.
One particular choice that does not break any translational symmetry follows from our construction of the low energy theory 
in Section~\ref{sec:continuum}: choose the condensate as in Eq.~(\ref{eq:rho0p}) and then all gauge invariant condensates that can be constructed
out of it are translationally invariant. In this case, we observe from Eq.~(\ref{eq:l1}) that there is a term linear in the monopole
operator $\psi$ in the action, and a consequent mixing of $\psi_\uparrow$ and $\psi_{\downarrow}^\dagger$.\\
({\em iv\/}) \underline{Paired boson supersolid}: Now we restore spin rotation invariance in the AFM
paired boson superfluid by condensing vortices in the spinons $z_\alpha$:
\beq
\langle \psi_\uparrow \rangle \neq 0~~,~~\langle \psi_\downarrow \rangle \neq 0~~,~~\langle \varphi_{+\ell} \rangle = 0~~,~~\langle \varphi_{- \ell} \rangle = 0.
\eeq
We assume the Cooper pair vortex bilinear condensates that are present here are the same as those in
the AFM paired boson superfluid. For the choice as in Eq.~(\ref{eq:rho00}), the present phase will
have VBS order, as in the insulator (see Fig.~\ref{fig:ins}). This is because now the combination $\rho_{00} = \psi_\uparrow \psi _\downarrow
\gamma_{00}$ has a non-zero expectation value, and this is a gauge-invariant observable which transforms
like the VBS order parameter (see Eq.~(\ref{defrho})). The transition to this VBS supersolid state from the AFM state in ({\em iii\/})
will be of the easy-plane CP$^1$ variety, just as in the insulator. The second possibility noted above
was to choose the Cooper pair vortex condensate as in Eq.~(\ref{eq:rho0p}): in this case the present phase
will {\em not\/} break translational symmetry, and will indeed be identical to the paired boson superfluid
in ({\em ii\/}). For the transition between the AFM paired boson superfluid and the paired boson superfluid
({\em i.e.\/} between the states in ({\em iii}) and ({\em ii})), the mixing between $\psi_\uparrow$ and $\psi^{\dagger}_\downarrow$ implies
that it is described by a single complex scalar (which is a linear combination of the these fields) coupled to the $a$ gauge field.
By Dasgupta-Halperin duality \cite{dh}, this transition is in the O(2) universality class (and in the O(3) class with full spin-rotation symmetry)---this is then nothing but the 
{\em conventional\/}
SDW transition.  We will consider other choices for the condensates of $\gamma_{mn}$ in Section~\ref{sec:max}
below, and also find one in which the paired boson superfluid is a nematic, for a case in which the density in the corresponding
state in ({\em iii\/}) has full square lattice symmetry. Other cases lead to a variety of supersolids
with density oscillation 
periods which are 
characteristic of the total boson density $1+x$.

\subsection{Loss of AFM order in the paired boson superfluid}
\label{sec:max}

This subsection deals with the nature of the superfluid phases ({\em iii\/}) and ({\em iv\/}) above, and of the 
quantum phase transition between them. Both phases have a paired boson condensate, but no single boson condensate,
and so the flux quantum is $h/(2e)$. The first phase also has antiferromagnetic (AFM) order, and spin rotation invariance
is restored in the transition to the second phase. We will see here that a rich variety of cases are possible, and we will present
a few illustrative examples.

We will argue that generically both the magnetically ordered and disordered phases break lattice symmetries. It is possible that the pattern of lattice symmetry breaking on the two sides of the magnetic phase transition is the same, in which case we expect a critical point in the O(3) universality class; this includes the case where there is no lattice symmetry breaking in either state,
as just noted above. It is also possible to have a phase transition where antiferromagnetism is lost, but a larger subgroup of the lattice symmetry is broken. Such transitions will either be first order or exotic (for instance, of a deconfined variety). We construct some specific examples of various scenarios. 

We would like to describe the paired boson phase, in which $\langle g_+ g_-\rangle \neq 0$. This operator corresponds to a monopole of the $b$ field, hence $b$ must not be Higgsed in this phase. On the other hand, we would like to suppress the single boson condensates $z_s g^{\dagger}_+$ and $z_s g_-$, which correspond to monopoles of flux $1/2$ in all three of the gauge fields. This can be achieved by Higgsing the gauge field $c$, which leaves the freedom for the gauge field $a$ to be either in the Higgs or Coulomb phase - i.e. we can consider the loss of antiferromagnetism in the presence of a paired boson condensate. 

We will Higgs $c$ by a condensate $\langle e^{i (\beta_+ - \beta_-)} \rangle \neq 0$. Note that this condensate is not charged under the $b$ field, and hence $b$ remains unhiggsed as desired. If we, instead, condensed $e^{i \beta_+}$ and $e^{i \beta_-}$, independently, then both $b$ and $c$ would be Higgsed and the resulting state would be an insulator.

We would now like to integrate the fluctuations of $\varphi$ and $c$ fields out, obtaining an effective theory for the spinon vortices $\psi_{\uparrow} = e^{i \alpha_{\uparrow}}$ and $\psi_{\downarrow} = e^{i \alpha_{\downarrow}}$ interacting with the gauge field $a$. In principle, the massless gauge field $b$, corresponding to the superfluid goldstone, also has to be included in the effective theory, however, as argued previously, it will decouple at low energies. The resulting theory will have two phases. In one phase, $\langle \psi_{\uparrow} \rangle = 0$, $\langle \psi_{\downarrow} \rangle = 0$ and $a$ is massless - this is the antiferromagnetic phase. In the other phase, the spinon vortices condense, $\langle \psi_{\uparrow}\rangle \neq 0$, $\langle \psi_{\downarrow} \rangle \neq 0$, the gauge field $a$ is Higgsed and antiferromagnetism is lost.

It is clear that for a generic set of condensates $\langle \gamma_{m n} \rangle$ the lattice symmetry is broken because we
can construct gauge invariant observables like $\gamma^{}_{mn} \gamma^\dagger_{m'n'}$ which transform non-trivially under
the space group symmetry. This has important consequences for the structure of the effective theory for $\psiu$, $\psid$ fields governing the loss of antiferromagnetism. The continuum action for these fields will have the form,
\beq L_{\psi} = L_0 + L_{m}\eeq
\beq L_0 = \frac{1}{2 \tilde{e}^2} (\nabla \times a)^2 +|(\d_{\mu} - 2 \pi i a_{\mu}) \psiu|^2 + |(\d_{\mu} + 2 \pi i a_{\mu}) \psid|^2 + \tilde{U}(|\psiu|^2, |\psid|^2)\eeq
where $\tilde{U}$ is a potential term invariant under $\psiu \leftrightarrow \psid$. 
 $L_0$ contains the lowest dimension operators invariant under independent phase rotations of $\psiu$ and $\psid$.\footnote{Here we've also assummed either direct or dual inversion symmetry to rule out single derivative terms. Moreover, we always assume time inversion symmetry.} However, due to the presence of the last term in (\ref{Sduallat}), or more physically, due to the compactness of the direct gauge field $A$, only the combination (\ref{U1a}) is a symmetry of the theory. Thus, we have an additional term $L_m$, generated by the monopoles of the direct theory, which will break the ``flux symmetry"
\beq  U(1)_{\Phi}:\,\,\psiu \to e^{i \theta/2} \psiu,\quad \psid \to e^{i \theta/2} \psid\eeq
The simplest terms in $L_m$ will be polynomials in the monopole operator $\psi = \psiu \psid$, which transforms as $\psi \to e^{i \theta} \psi$ under $U(1)_{\Phi}$. 		 
In the well understood case of a pure spin system, lattice symmetry (\ref{PSGmon}) implies that only ``quadrupled monopoles" survive and the lowest order term allowed in $L_m$ is,
 \beq L_m = -y_m (\psi^4 + (\psi^{\dagger})^4)\label{psi4}\eeq
Intuitively, this quadrupling is due to the presence of oscillating Berry phases $\zeta_j$ in (\ref{Sduallat}), which lead to a destructive interference of single monopoles. We see that the action (\ref{psi4}) preserves a ${\mathbb Z}_4$ subgroup of $U(1)_\Phi$, which from the transformation in Eq.~ (\ref{PSGmon}) is identified with (direct) lattice rotations. 
More generally, as noted in Section \ref{sec:dual}, $\psi$ has the transformation properties of a valence-bond-solid (VBS) order parameter. 
On the magnetically ordered side of the phase diagram, the ${\mathbb Z}_4$ symmetry is unbroken, $\langle \psi \rangle = 0$, and hence the lattice symmetries are preserved, while the magnetically disordered side breaks the ${\mathbb Z}_4$ symmetry via, $\langle \psi \rangle = \langle \psiu \psid \rangle \neq 0$ leading to a VBS order. The term (\ref{psi4}) is expected to be irrelevant at the critical point where the magnetic order is lost and hence monopoles are suppressed at the phase transition. Undualizing back to the direct theory, we obtain a model where the gauge-field $A$ is non-compact and the resulting critical point is of a ``deconfined" variety. In particular, a direct second order transition between the two phases is allowed.

The above picture is still expected to hold in the present model at zero doping in the presence of a paired doublon superfluid, as the condensate $\langle e^{i (\beta_+ - \beta_-)}\rangle$ does not break any lattice symmetries. However, once we go to doublon superfluid states at finite doping and develop condensates $\langle \gamma_{mn}\rangle$ - we generically break lattice symmetry. Hence, lattice symmetry will be broken both in the magnetically ordered and disordered phases. Moreover, the monopole term in the spinon vortex action is no-longer constrained by Eq.~(\ref{PSGmon}), and single monopole terms will be generated from the coupling in Eq.~(\ref{eq:l1}),
\beq L_m = -(y_m \psi + y^*_m \psi^{\dagger}). \label{psi1}\eeq
The fact that the monopoles are no longer quadrupled is roughly due to the spatially oscillating nature of the condensate $\langle e^{i (\beta_+ - \beta_-)}\rangle$, which cancels the Berry phases in eq. (\ref{Sduallat}). As we know, the direct gauge theory with monopoles allowed is equivalent to the O(3) $\sigma$-model (or its easy plane counterpart in the present case). Thus, we will have a phase transition in the O(3) universality class (or O(2) class in the easy plane case). This is consistent with our expectations since only the Neel order is lost and no new lattice order is gained across the phase transition.

We note that though the above scenario is the most general one at finite doping, it is possible that the set of non-zero condensates $\langle \gamma_{mn}\rangle$ does not break the lattice symmetry, or break it only partially. Note that for a lattice operator $g$ to be unbroken, it is enough that the product of $g$ and a gauge rotation be preserved. So if the lattice symmetry is preserved by $\langle \gamma_{mn}\rangle$ up to a rotation in the gauge group $U(1)_c$, the transformations of the monopole field $\psi$ under the unbroken symmetries might be modified from the ones at zero doping in eq. (\ref{PSGmon}). This unleashes a whole set of different possibilities for phase transitions out of the antiferromagnetic phase, accompanied by breaking of additional parts of the lattice symmetry. We will present some examples of this scenario below.
 

However, first we would like to discuss an alternative way of looking at the magnetically restored phase, where the spinon vortices $\psiu$, $\psid$ are condensed. 
 So far, we have been thinking about the way the condensation of vortex-antivortex 
 pairs $\langle \varphi^{\dagger}_{-l}\varphi_{+n}\rangle$ affects the spinon vortices. This is the correct logic for studying the phase transition where antiferromagnetism is lost. However, once the spinon vortices (or more generally the monopole field $\psi$) are condensed, it is instructive to ask the reverse question: how are the Cooper pair vortices affected? The condensates $\psiu$, $\psid$ Higgs the $c$ (and $a$) gauge fields and appear to break lattice symmetries. However, as long as $|\langle \psiu\rangle| = |\langle \psid\rangle|$ (i.e. time reflection symmetry is unbroken), a combination of lattice and $U(1)_c$ rotations is always preserved. In particular, by a gauge rotation we can choose $\psiu = \psid$ real, and then $T_x (i)_c$, $T_y$, $R^{\mathrm{dual}}_{\pi/2}(e^{i \pi/4})_c$, $I_x$, ${\cal T}$ are preserved. Under these symmetries, the Cooper pair vortices transform as (we list only translations here for brevity),
\bea && \bar{T}_x = T_x(i)_c :\,\,  \varphi_{+l} \to i \varphi_{-,{l+1}}, \quad \varphi_{-l} \to -i \varphi_{+,l+1}\nn\\
		 && \bar{T}_y = T_y :\,\,  \varphi_{+l} \to \omega^{-l} \varphi_{- l},\quad \varphi_{-l} \to \omega^{-l} \varphi_{+l}\eea
We observe,
\beq \bar{T}_y \bar{T}_x = (\bar{\omega})_b \bar{T}_x \bar{T}_y\eeq
with,
\beq \bar{\omega} = - \omega = e^{2 \pi i \bar{p}/\bar{q}}\eeq
where
\beq \frac{\bar{p}}{\bar{q}} = \frac{1+x}{2}\eeq
Hence, the condensed monopoles endow the Cooper pair vortices with a new projective implementation of the lattice symmetry. Comparing with eq. (\ref{dopingpq}) we conclude that the condensed monopoles shift the effective density of boson from $x$ to $1+x$, as we claimed above.

Now, we come back to specific examples of phase transitions out of the antiferromagnet. In Sections~\ref{sec:gamma00}, \ref{sec:gammap0},
and \ref{sec:gammapp} we will provide a complete listing of AFM paired boson superfluids which have full square lattice symmetry
in the density; these cases will lead to transitions to the paired boson superfluid as shown in Fig.~\ref{fig:phases2}. The remaining
Sections~\ref{sec:gammap2}, \ref{sec:gammap3} will consider cases in which square lattice symmetry is broken in both 
the AFM and non-magnetic phases, but with the nature of the lattice symmetry breaking changing across the transitions.

\subsubsection{$\langle \gamma_{00}\rangle \neq 0$: Deconfined critical point to a valence-bond-solid ($x$ - arbitrary)}
\label{sec:gamma00}

\begin{figure}[t]
\includegraphics[width=1.75in]{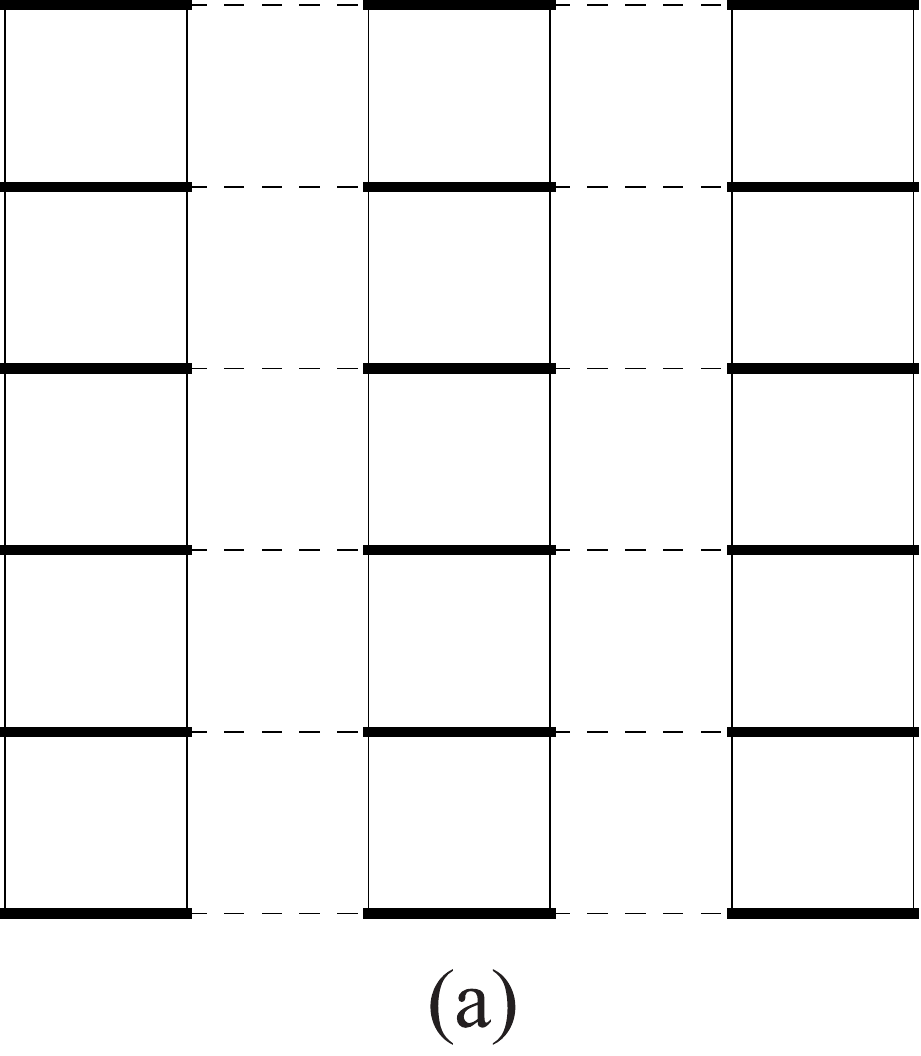}\quad\quad\quad\quad\includegraphics[width=1.75in]{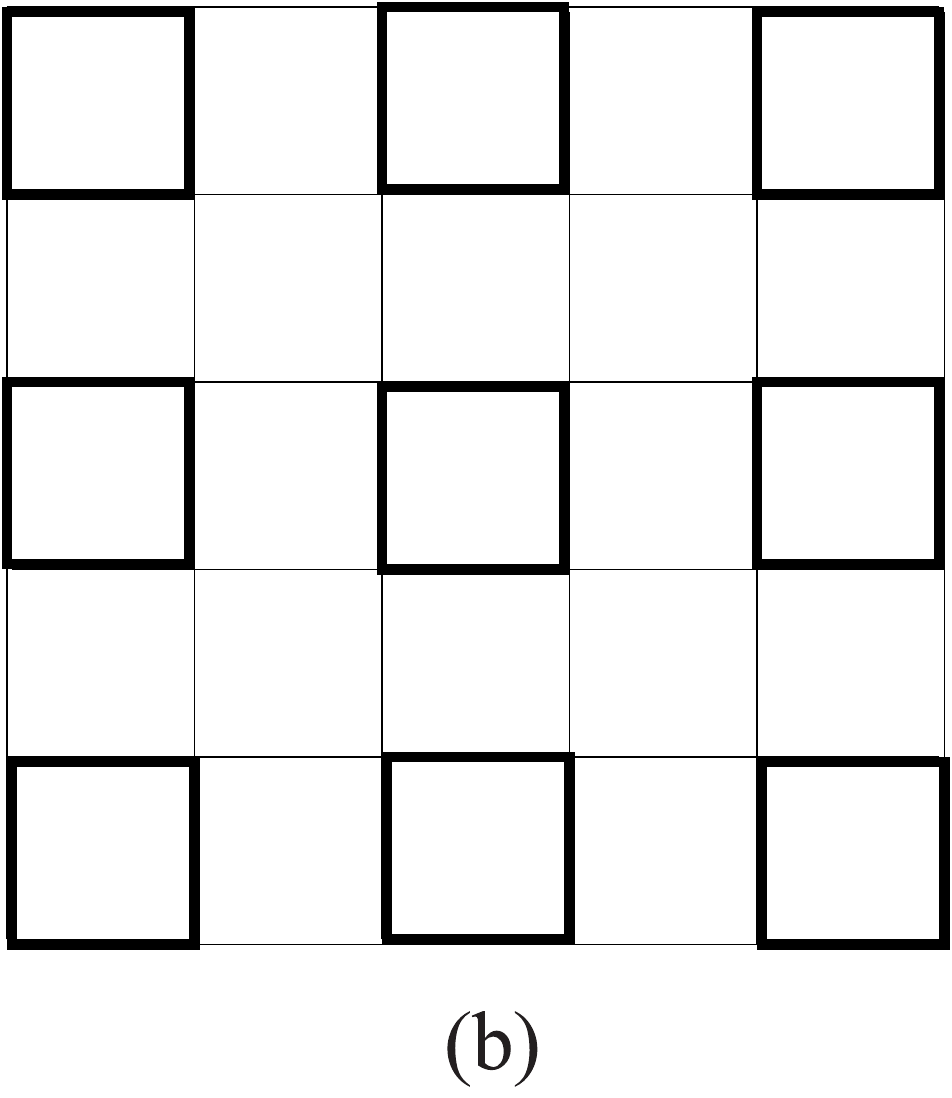}
\caption{Orderings in the non-magnetic state in the presence of the $\langle \gamma_{0 0}\rangle$ condensate (all $q$),
or the $\langle \gamma_{\pi \pi}\rangle$ condensate ($q/2$ even).}
\label{fig:ins}
\end{figure}

We imagine that the vortex-antivortex condensate present is $\langle \gamma_{00}\rangle$. As already noted in Section \ref{sec:dual}, this condensate does not break any lattice symmetries. Hence, the discussion given for zero doping applies here. Namely, the antiferromagnetic state will break no lattice symmetries, and we will have a transition to a valence-bond-solid state via a deconfined quantum critical point described by the CP$^1$ field theory. 
The possible patterns of square lattice symmetry 
breaking are as in the insulator, and illustrated in Fig.~\ref{fig:ins}.

\subsubsection{$\langle \gamma_{\pi 0}\rangle,\,\langle \gamma_{0 \pi}\rangle \neq 0$ ($q$ - even)}
\label{sec:gammap0}

The simple example below illustrates the range of possibilities for the phase transition out of a paired doublon antiferromagnet.

We assume that $q$ is even and $\langle \gamma_{\pi 0}\rangle$, $\langle \gamma_{0 \pi}\rangle$ are non-zero. However, a particular choice of $\langle \gamma_{\pi 0}\rangle$, $\langle \gamma_{0 \pi}\rangle$ exists where by combining lattice operations with $U(1)_c$ transformations, all lattice symmetries can be preserved. The resulting symmetries are $T_x(i)_c$, $T_y$, $I^{\mathrm{dual}}_x$, ${\cal T}$ and $R^{\mathrm{dual}}_{\pi/2}(e^{\pm i \pi/4})_c$
(we will drop transformations under the time reflection symmetry below, since it is never broken by $\langle \gamma_{mn} \rangle$). As we will see the factors $(e^{\pm i \pi/4})_c$ for $R^{\mathrm{dual}}_{\pi/2}$ corresponding to the presence of condensates $\langle(\gamma_{\pi 0} \mp i \gamma_{0 \pi})\rangle$ lead to two inequivalent scenarios. The new transformations for the monopole operator $\psi$, therefore, are
\bea 
T_x: \,\,&&\psi \to \psi^{\dagger}\nn\\
T_y: \,\,&& \psi \to \psi^{\dagger}\nn\\
R^{\mathrm{dual}}_{\pi/2}:\,\,&& \psi \to \pm \psi^{\dagger}\nn\\
I^{\mathrm{dual}}_x: \,\,&& \psi \to \psi\label{PSGq2}\eea
Note the two different possible transformations under $R^{\mathrm{dual}}_{\pi/2}$. If we choose the $+$ sign, then the Berry phases of spinon vortices and Cooper pair vortices cancel each other, and the lowest allowed term in $L_m$ is,
\beq L_m = -y_m (\psi + \psi^{\dagger}) ;\eeq
We can also view this term as arising from Eq.~(\ref{eq:l1}).
Thus, we obtain a theory with unsuppressed monopoles, and expect a phase transition in the O(3) (O(2) in the present easy plane model) universality class. Note that the lattice symmetry will be unbroken on both sides of the phase transition. This is the conventional SDW transition between AFM paired boson superfluid and paired boson superfluid discussed in Section \ref{sec:dual}.

\begin{figure}[t]
\includegraphics[width=1.75in]{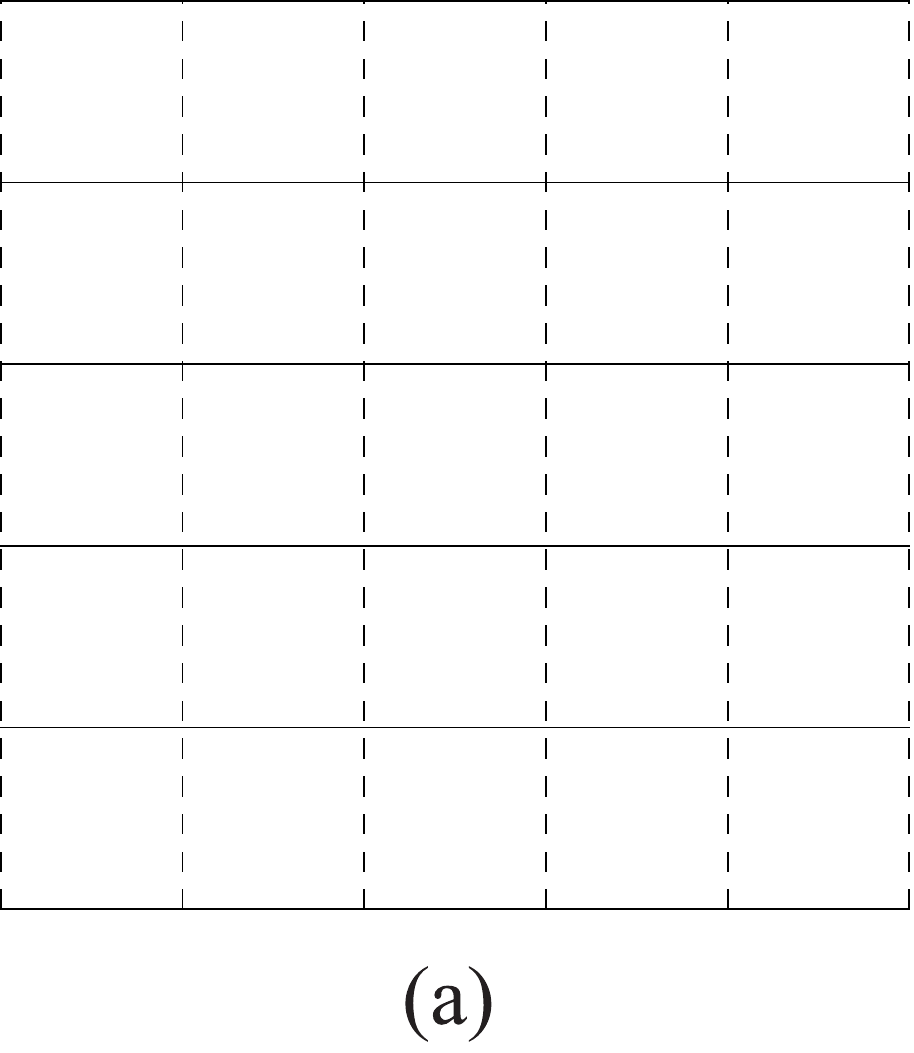}\quad\quad\quad\quad\includegraphics[width=1.75in]{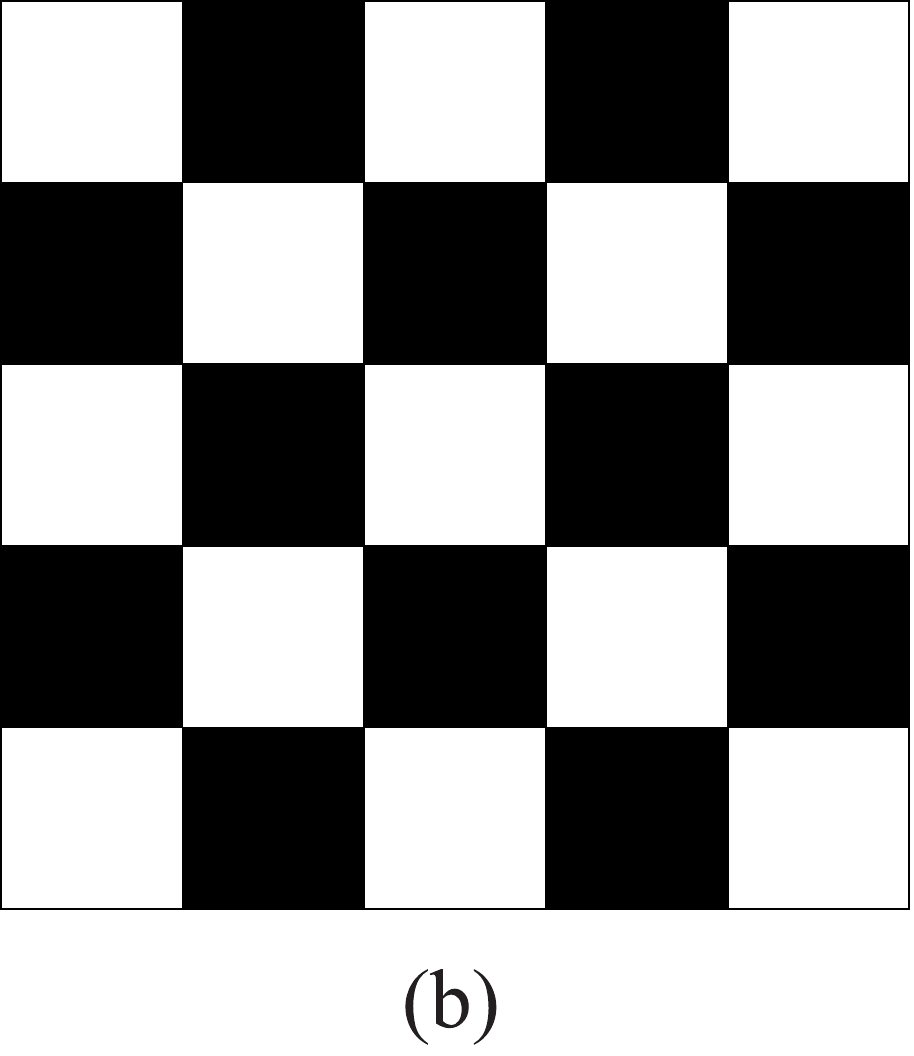}
\caption{Ordering in the non-magnetic state in the presence of $\langle \gamma_{\pi 0}\rangle$, $\langle \gamma_{0 \pi}\rangle$ condensates.}
\label{q2}
\end{figure}

Alternatively, if we choose the $-$ sign for $R^{\mathrm{dual}}_{\pi/2}$, the Berry phases of $\psi$'s and $\varphi$'s add-up and Eq.~(\ref{PSGq2}) become the transformations of a monopole operator in an antiferromagnet with odd-integer spins. The lowest allowed term in $L_m$ is,
\beq L_m = -y_m (\psi^2 + (\psi^{\dagger})^2)\label{doubledmon}\eeq
and the monopoles are ``doubled"; this term arises from Eq.~(\ref{eq:l2}). The residual ${\mathbb Z}_2$ flux symmetry corresponds to direct lattice rotations. Thus, in the antiferromagnetic phase $\langle \psi \rangle = 0$ and the lattice symmetry is unbroken. In the non-magnetic phase, $\langle \psi \rangle \neq 0$ and lattice symmetry is broken - there are two different patterns for this depending on microscopic details (similar to dimer and plaquette states of a VBS). In one case, $\langle \psi \rangle = \langle \psi^{\dagger} \rangle$ and the only broken symmetry is $R^{\mathrm{dir}}_{\pi/2}$ (broken to $(R^{\mathrm{dir}}_{\pi/2})^2$). This is the nematic
superconductor, and a schematic picture of this state is given in Fig. \ref{q2}a. In the other case, $\langle \psi \rangle = - \langle \psi^{\dagger} \rangle$ and the lattice symmetry is broken to $T_x T_y$, $T_x T^{-1}_y$, $R^{\mathrm{dual}}_{\pi/2}$, $I^{\mathrm{dual}}_x$. A schematic picture of this state is given in Fig. \ref{q2}b. As for the nature of the phase transition in this case, it is expected that the doubled monopole is a relevant operator, which can lead to a direct first order phase transition.

\subsubsection{$\langle \gamma_{\pi \pi}\rangle \neq 0$ ($q$ - even)}
\label{sec:gammapp}

This is another case in which the antiferromagnetic superconductor has no density wave order.
The nature of the non-magnetic supersolid requires a separate analysis for $q/2$ even and odd.

For $q/2$ even, we observe from Eqs.~(\ref{defgamma}) and (\ref{PSGrho}), in a gauge 
where $\langle \gamma_{\pi \pi}\rangle$ is purely imaginary,
the square lattice symmetry operations are $T_x$, $T_y$, $I^{\mathrm{dual}}_x$, and $R^{\mathrm{dual}}_{\pi/2}(i)_c$.
The new transformations for the monopole operator $\psi$, therefore, are
\bea 
T_x: \,\,&&\psi \to -\psi^{\dagger}\nn\\
T_y: \,\,&& \psi \to \psi^{\dagger}\nn\\
R^{\mathrm{dual}}_{\pi/2}:\,\,&& \psi \to -i \psi^{\dagger}\nn\\
I^{\mathrm{dual}}_x: \,\,&& \psi \to \psi\eea
It is now easily seen that this case is the same as the supersolid for $\langle \gamma_{00} \rangle \neq 0$ above,
and the VBS order is as in Fig.~\ref{fig:ins}. The transition between the AFM and non-magnetic states
is described by the deconfined CP$^1$ theory.

\begin{figure}[t]
\includegraphics[width=1.75in]{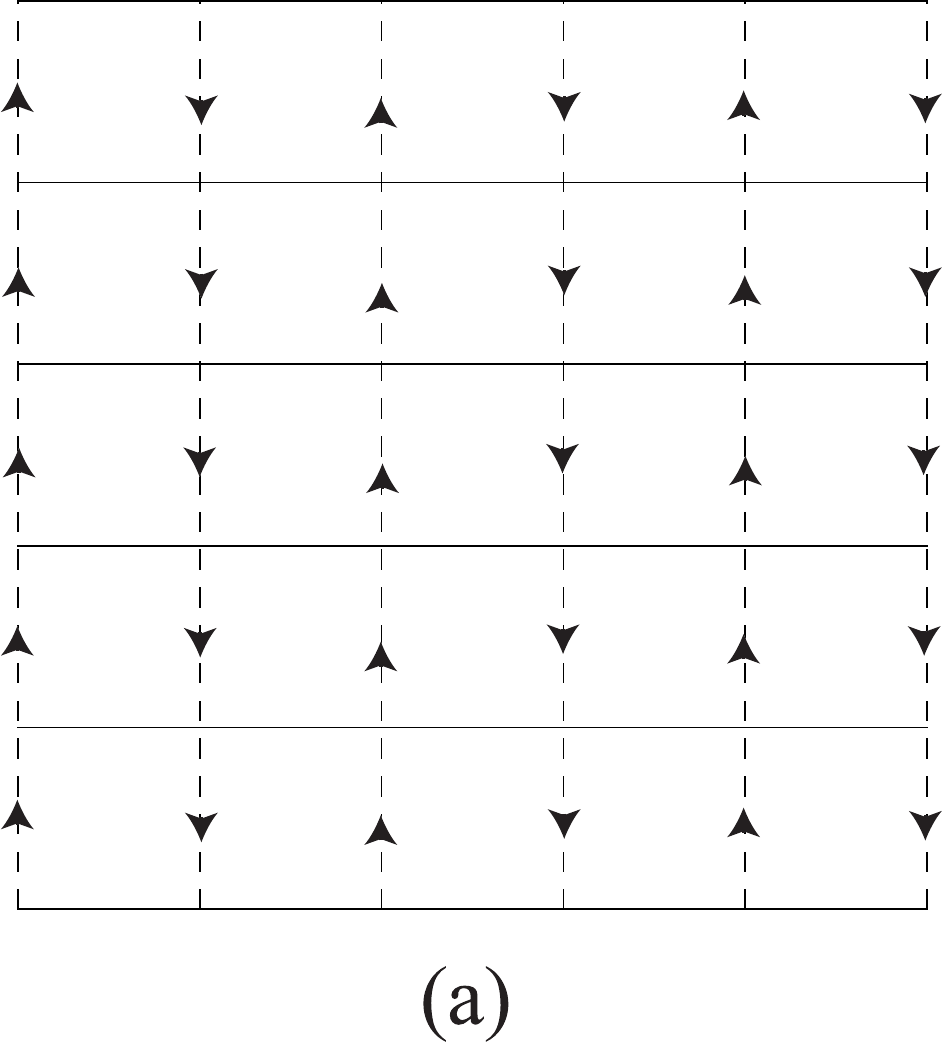}\quad\quad\quad\quad\includegraphics[width=1.75in]{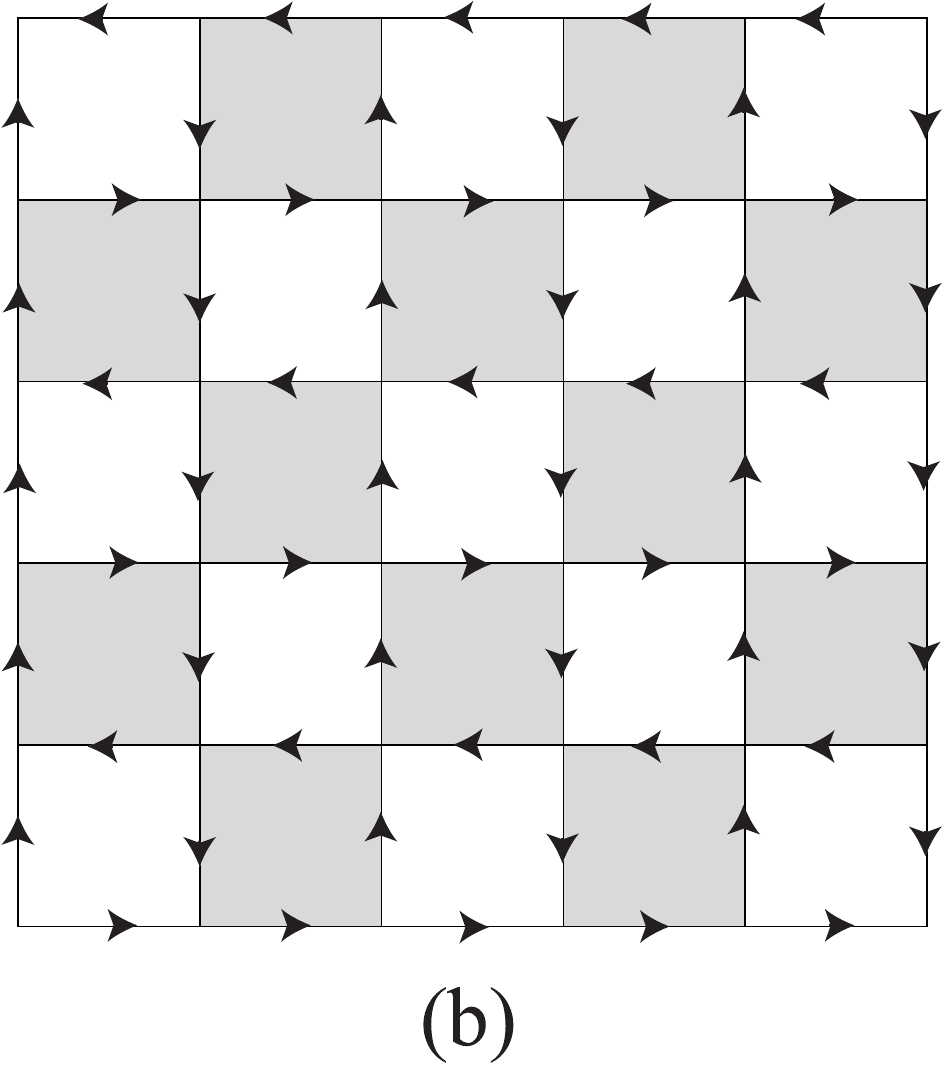}
\caption{Ordering in the non-magnetic state in the presence of $\langle \gamma_{\pi \pi}\rangle$ condensate for $q/2$ odd.
The states are defined by Eq.~(\ref{eq:q2ab}) for (a) $\arg(\langle \psi \rangle) = 0$ and (b) $\arg(\langle \psi \rangle) = \pi/4$.
Both states are 4-fold degenerate.
The line patterns indicate the link energy variable $Q_{ij}=\vec{S}_i \cdot \vec{S}_j$, while the arrows indicate
the directed link variable $P_{ij} = (\vec{S}_i \cdot \vec{S}_j )( \vec{S}_i^2 - \vec{S}_j^2 ) $. Note that the arrows do not imply
spin or charge currents (time reversal symmetry is preserved), 
and the state can be fully characterized by modulations in the charge density along the links.
}
\label{fig:q2ab}
\end{figure}

For $q/2$ odd, we find from Eqs.~(\ref{defgamma}) and (\ref{PSGrho}), again in a gauge 
where $\langle \gamma_{\pi \pi}\rangle$ is purely imaginary,
the square lattice symmetry operations of the antiferromagnet 
are $T_x$, $T_y$, $I^{\mathrm{dual}}_x (i)_c$, and $R^{\mathrm{dual}}_{\pi/2}$, and
therefore
\bea 
T_x: \,\,&&\psi \to -\psi^{\dagger}\nn\\
T_y: \,\,&& \psi \to \psi^{\dagger}\nn\\
R^{\mathrm{dual}}_{\pi/2}:\,\,&& \psi \to i \psi^{\dagger}\nn\\
I^{\mathrm{dual}}_x: \,\,&& \psi \to - \psi \label{eq:q2ab} \eea
Again, only quadrupled monopoles as in Eq. (\ref{psi4}) are allowed. Now in a non-magnetic state with $\langle \psi \rangle \neq 0$, as above, the symmetry of the state
is different for $\arg (\langle \psi \rangle) = 0, \pi/2, \pi, 3\pi/2$ and $\arg (\langle \psi \rangle) = \pi/4, 3\pi/4, 5\pi/4, 7\pi/4$.
For the first case, the state with $\arg(\langle \psi \rangle) = 0$ preserves $T^2_x$, $T_y$, $(R^{\rm dual}_{\pi/2})^2$, and  $I^{\rm dir}_x$,
while for the second case, the state with $\arg(\langle \psi \rangle) = \pi/4$ preserves $T^2_x$, $T^2_y$, $R^{\rm dual}_{\pi/2}$, and
$I^{\rm dual}_x T_x T_y$. Unlike other states we have considered here, these states cannot be constructed purely out of modulations of the
bond energy variable $Q_{ij} = \vec{S}_i \cdot \vec{S}_j$, where $\vec{S}_i$ is the spin operator on site $i$ of the direct square lattice.
Instead, we also need a {\em directed} bond variable $P_{ij} = (\vec{S}_i \cdot \vec{S}_j )( \vec{S}_i^2 - \vec{S}_j^2 ) $ which is a spin singlet
observable obeying $P_{ij} = - P_{ji}$. Note that because we are considering a doped system, the on-site spin fluctuates between
$\vec{S}_i^2 = 0,3/4$, and so $P_{ij}$ is not identically zero. The spatial modulations in these variables in the states for $q/2$ odd are shown in 
Fig.~\ref{fig:q2ab}.
As was the case for $q/2$ even, the transition between the AFM and non-magnetic states
is described by the deconfined CP$^1$ theory.

\subsubsection{$\langle \gamma_{\pm{\pi/2},0}\rangle \neq 0$ 
($q \equiv 0 \, (mod \,4)$)}
\label{sec:gammap2}

Here we consider a transition in the background of condensates $\langle \gamma_{\pm \pi/2,0} \rangle$ for $q$ divisible by $4$.
As we will see, one of the possible magnetically disordered states in this case is a charge density wave with period $4$.
Such a state is actually observed on the hole-doped side of the cuprate phase diagram close to the doping $x = 1/8$ ($p=1$, $q = 16$)\cite{kohsaka}. We note, however, that we reach such a state only from an AFM superconductor which already has density wave order
as in Fig.~\ref{fig:ins}a.

\begin{figure}[t]
\includegraphics[width=1.75in]{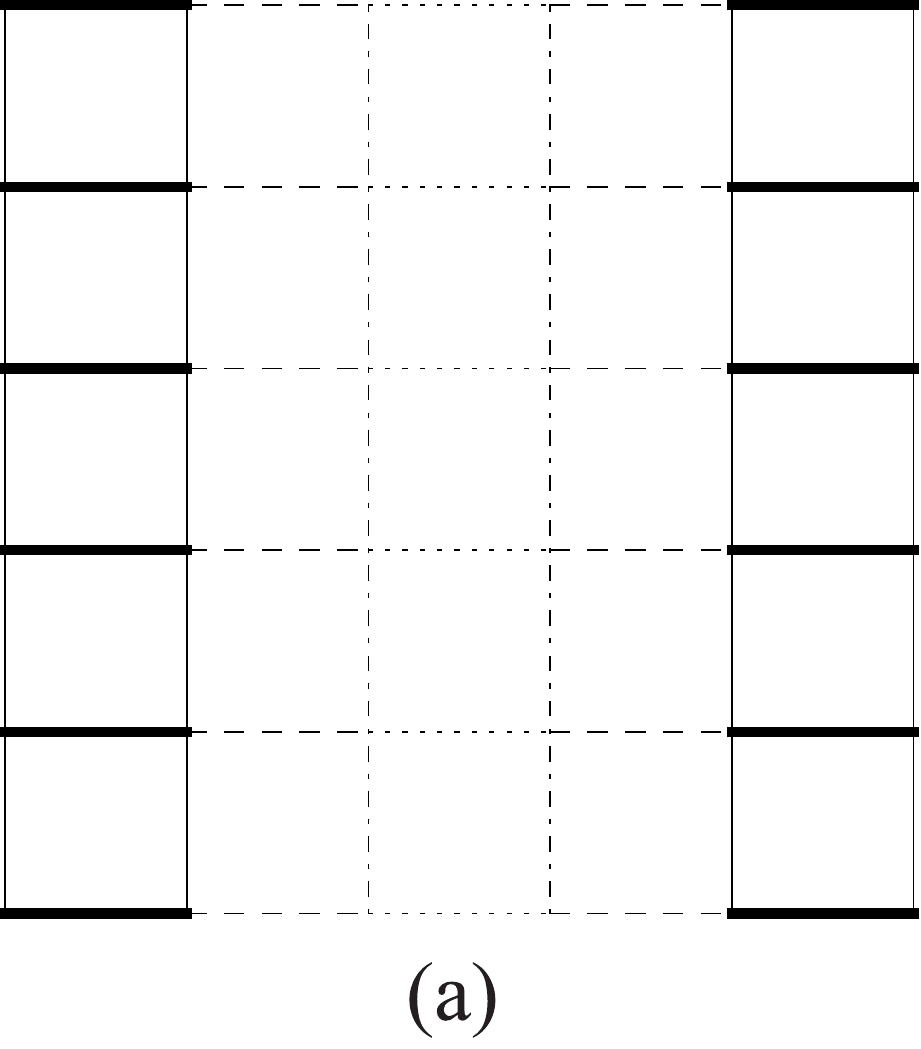}\quad\quad\quad
\includegraphics[width=1.75in]{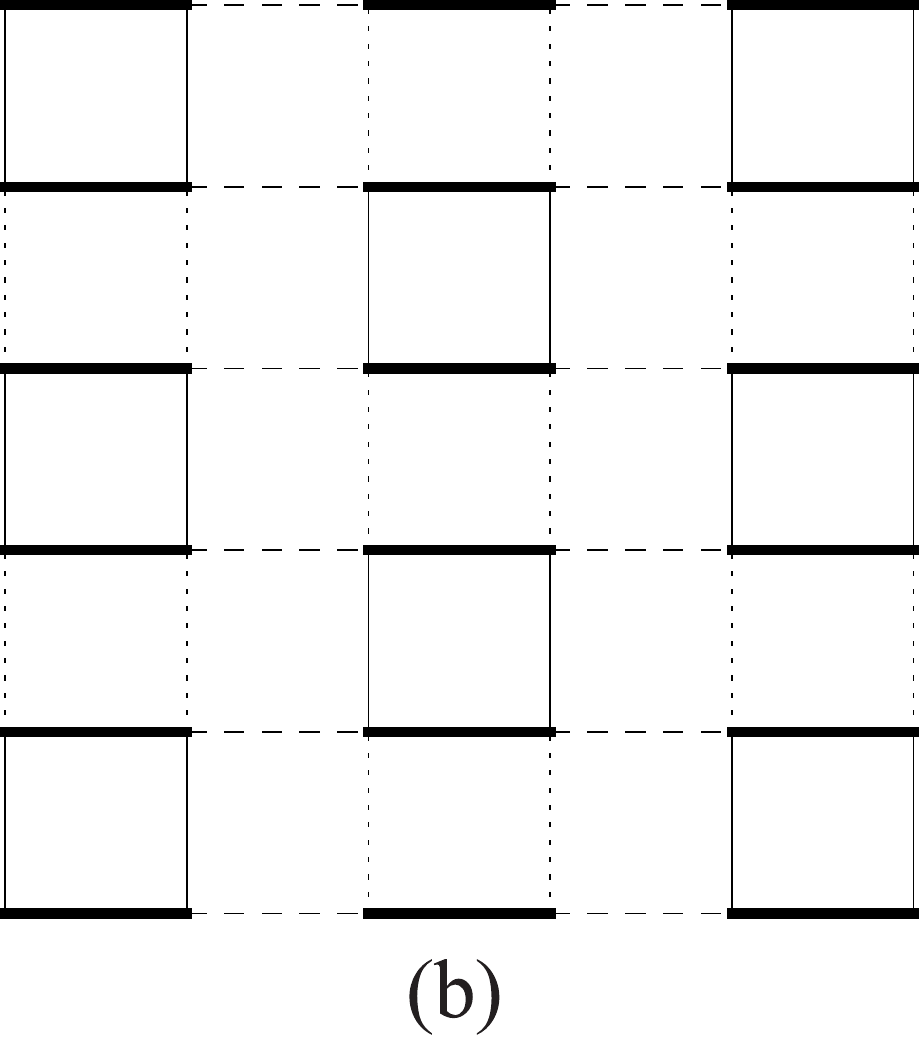}
\caption{An example of a phase transition out of an antiferromagnet with dimer order as in Fig.~\ref{fig:ins}a, to the non-magnetic states (a) and (b)
above.}
\label{q4}
\end{figure}

It turns out that once the condensates $\langle \gamma_{\pm \pi/2,0} \rangle$ are present, $T_x$ is automatically broken, as can be seen from the transformation properties of the gauge invariant observable $\langle \gamma^{}_{\pi/2,0} \gamma^{\dagger}_{- \pi/2,0}\rangle$. However, one can still arrange
for $T_y$ and $T^2_x(i)_c$ to be preserved (generally, the condensates $\langle \gamma_{\pm\pi/2,\pi} \rangle$ will also be allowed by these symmetries). Moreover, we have the choice of preserving rotations by $180^\circ$ about either direct or dual lattice site. For brevity we will only discuss the later case, as it might be physically relevant. Then one can preserve $T^2_x(i)_c$, $T_y$, $(R^{\mathrm{dual}}_{\pi/2})^2$ and $I^{\mathrm{dual}}_x$. We recognize that a state with such a symmetry is a valence-bond-solid, see Fig. \ref{fig:ins}a, with superposed antiferromagnetic order. The transformation properties of the monopole operator under the remaining symmetry group are,
\bea T^2_x : \,\, && \psi \to - \psi\nn\\
T_y :\,\, && \psi \to \psi^{\dagger}\nn\\
(R^{\mathrm{dual}}_{\pi/2})^2: \,\, && \psi \to \psi\nn\\
I^{\mathrm{dual}}_x: \,\, && \psi \to \psi\eea
Thus, we again have the case of ``doubled monopoles," eq. (\ref{doubledmon}).
Note, however, that now the residual ${\mathbb Z}_2$ flux symmetry corresponds to translations by two lattice sites along the $x$ direction.
So the antiferromagnet has $\langle \psi \rangle = 0$ and carries a VBS order. Once antiferromagnetism is lost and $\langle \psi \rangle \neq 0$, we break an additional subgroup of the lattice symmetry. There are again two cases:\\
({\em a\/}) $\langle \psi \rangle = \langle \psi^{\dagger} \rangle$. Then the remaining symmetry group is $T^4_x$, $T_y$, $(R^{\mathrm{dual}}_{\pi/2})^2$ and $I^{\mathrm{dual}}_x$. This state is a bond centered charge-density-wave with period four. A cartoon picture of this state is shown in Fig. \ref{q4}a. Precisely such a configuration is observed by STM experiments on hole doped cuprates near $x = 1/8$. \cite{kohsaka}\\
({\em b\/}) $\langle \psi \rangle = - \langle \psi^{\dagger} \rangle$. Then the remaining symmetry group is $T^2_x T_y$, $T^2_x T^{-1}_y$,  $(R^{\mathrm{dual}}_{\pi/2})^2$ and $I^{\mathrm{dual}}_x$. A schematic picture of this state is shown in Fig. \ref{q4}b. 

Note that in both of the cases ({\em a\/}) and ({\em b\/}) we have a transition from a state with a unit cell area of $2$ to a unit cell area of $4$.

\subsubsection{$\langle \gamma_{\pm{\pi/2},0}\rangle \neq 0$ and $\langle \gamma_{0,\pm{\pi/2}}\rangle \neq 0$ ($q \equiv 0 \, (mod \,4)$).}
\label{sec:gammap3}

The present example is of interest as we will be able to construct a deconfined critical point other than the one separating
an antiferromagnet and a valence bond solid.

\begin{figure}[t]
\includegraphics[width=2in]{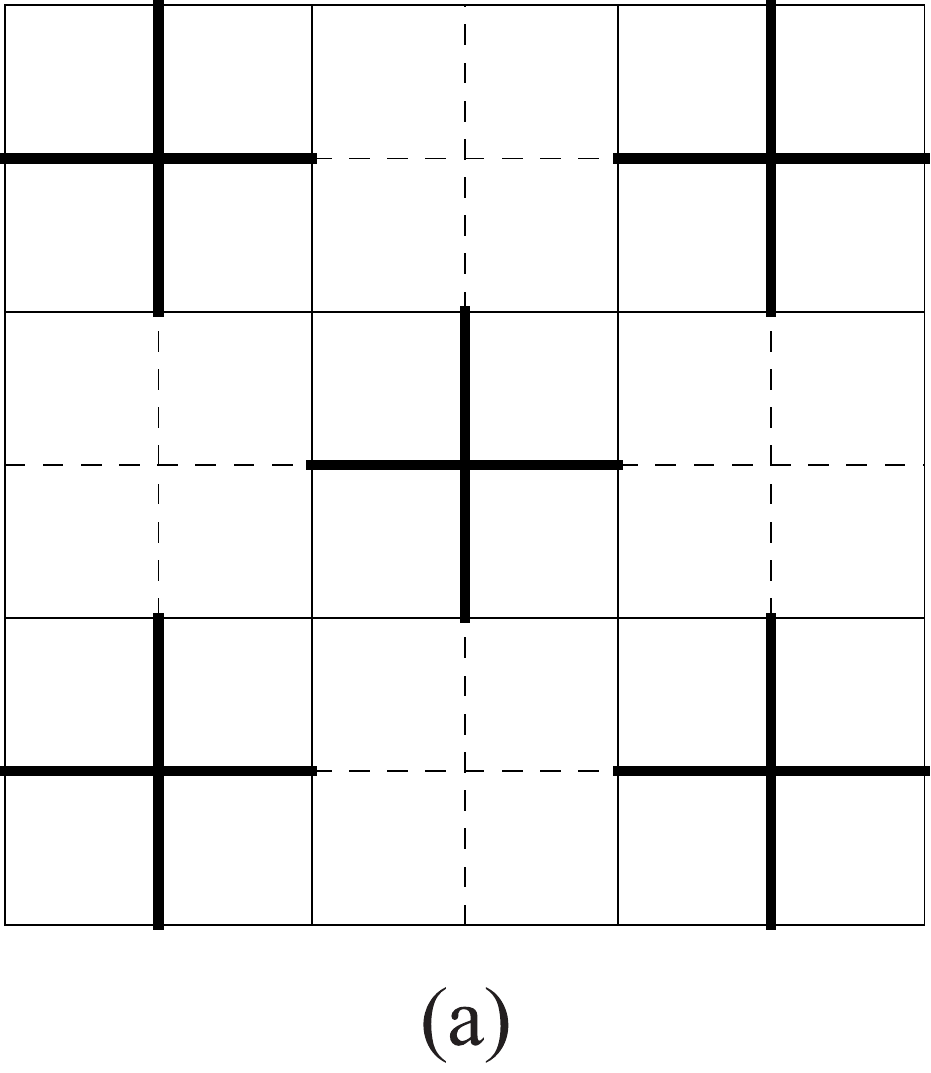}\quad\quad\quad\quad \includegraphics[width=2in]{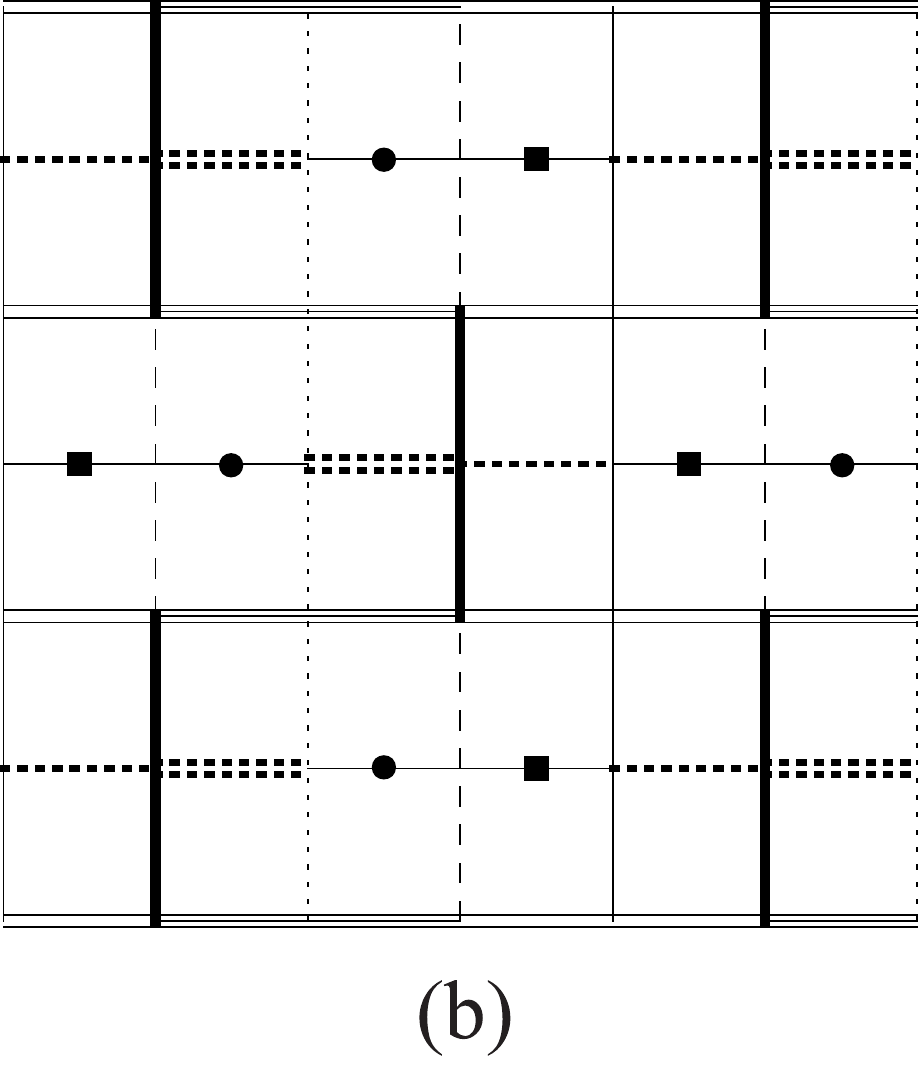}
\caption{A candidate for a new deconfined phase transition from magnetic state (a) to a non-magnetic state (b) with a higher degree of lattice
symmetry breaking.}
\label{figdir}
\end{figure}

Once $\langle \gamma_{\pm \pi/2,0}\rangle$ and $\langle \gamma_{0,\pm \pi/2}\rangle$ are turned on, both translations along $x$ and $y$ are broken. However, the combinations $T^2_x T^2_y (i)_c$ and $T^2_x T^{-2}_y (i)_c$ are preserved. Moreover, one can now arrange for rotations about either direct or dual lattice site to be preserved. We will concentrate on the direct lattice site case as it yields a deconfined critical point. Then, it turns out that we can mantain $I^{\mathrm{dir}}_x$ and either ({\em a\/}) $R^{\mathrm{dir}}_{\pi/2}$ or ({\em b\/}) $R^{\mathrm{dir}}_{\pi/2} (e^{-i\pi/4})_c$. The spatial modulations in such a state are shown schematically in Fig. \ref{figdir}a. The transformations of the monopole operator in the case ({\em a\/}) are,
\bea T^2_x T^2_y : \,\, && \psi \to - \psi\nn\\
T^2_x T^{-2}_y :\,\, && \psi \to -\psi\nn\\
R^{\mathrm{dir}}_{\pi/2}: \,\, && \psi \to i \psi\nn\\
I^{\mathrm{dir}}_x: \,\, && \psi \to -\psi^{\dagger}
\eea
with the case ({\em b\/}) differing only in the transformation under $R^{\mathrm{dir}}_{\pi/2}$,
\beq R^{\mathrm{dir}}_{\pi/2}: \,\, \psi \to \psi\eeq
Therefore, ``doubled" monopoles are permitted in case ({\em b\/}), making it of less theoretical interest. Below, we will, therefore, concentrate on the case ({\em a\/}). Here, as a consequence of a ${\mathbb Z}_4$ flux symmetry associated with direct lattice rotations, only ``quadrupled" monopoles are permitted as in Eq. (\ref{psi4}), and a deconfined phase transition may be possible. We note that we have addressed only the instability of the phase transition to monopole proliferation here. However, since the symmetry group in the present case is smaller than for the usual antiferromagnet - valence-bond-solid deconfined critical point, other relevant operators can arise, such as e.g. $(\epsilon_{\tau \mu \nu} \d_{\mu} a_{\nu}) (|\psi_{\uparrow}|^2 - |\psi_{\downarrow}|^2)$. A classification and RG treatment of such operators is beyond the scope of this work.

As for the pattern of spatial modulations: in the antiferromagnetic phase, we have $\langle \psi \rangle = 0$ and we obtain the state in Fig. \ref{figdir}a. Once antiferromagnetism is lost and $\langle \psi \rangle \neq 0$, we break an additional lattice subgroup. As for the usual deconfined critical point there are two possible cases: $\langle \psi \rangle \sim 1$ and $\langle \psi \rangle \sim e^{i \pi/4}$. In the first case, the remaining symmetries are $T^4_x$, $T^4_y$, $(R^{\mathrm{dir}}_{\pi/2})^2 T^2_x T^2_y$, $I^{\mathrm{dir}}_y$, and we get the state in Fig. \ref{figdir}b. In the second case, the remaining symmetries are $T^4_x$, $T^4_y$, $(R^{\mathrm{dir}}_{\pi/2})^2 T^2_x T^2_y$, $I^{\mathrm{dir}}_x R^{\mathrm{dir}}_{\pi/2}$ (for brevity we omit a figure of this state).

\section{Discussion}
\label{sec:conc}

We have discussed different possibilities for the destruction of N\'eel order in metallic or superconducting two-dimensional quantum antiferromagnets by doping in a small density of charge carriers into the parent insulators. We have summarized our results already in detail in the introduction, and so we will be brief here.

The standard SDW theory for the appearance of N\'eel order in a metal, generically requires an intermediate state between the large Fermi surface metal at overdoping and the small Fermi pocket state at very low doping.
This intermediate state has 8 zero crossings in the fermion dispersion along the Brillouin zone diagonals.
Because such an intermediate state has not so far been observed,
we have examined other routes to connecting such states.
We used a formalism which decomposed the electron operator as a product of bosonic spinons and fermionic spinless 
doublons. Despite our use of this `fractionalized' approach, one of our results was the remarkable reappearance
in this formalism of the conventional SDW criticality for the loss of N\'eel order in the superconductor. In addition, 
we also found other universality classes for the loss of magnetic order in the AFM+SC state which mimic those
found in insulating antiferromagnets for different values of $S$, as shown in Fig.~\ref{fig:phases2}. For the metallic case,
we found a transition to an algebraic charge liquid - the doublon metal.

Our endeavour was motivated by the fairly strong evidence for a magnetic
quantum critical point at which N\'eel order is lost in the electron-doped cuprates~\cite{greene04,greene07a,greene07b,greene07c,greven07,dai07}.
We hope that the scenarios presented here will be tested in future experiments. A clear strategy to do so has been provided in Sec.~\ref{sec:expts}.

In principle, our results here can be extended to the case of the hole-doped cuprates, which were considered earlier
in Refs.~\onlinecite{ffliq,acl}. The main phenomenological difficulty, as we noted in Section~\ref{sec:intro}, is that the antiferromagnetism
in the La based hole-doped cuprates does not remain pinned at $(\pi, \pi)$. 
However it may well be that $(\pi, \pi)$ antiferromagnetism is more important in the other hole-doped
cuprates. So, we can consider the transition from the antiferromagnetic metal
to the holon metal---all of our analysis here on the transition to the doublon metal carries over, and the transition is in the O(4) class.
Unlike the doublon superconductor, the holon superconductor is not immediately unstable to confinement.
The holon superconductor has $N_f = 4$ gapless Dirac fermion excitations which carry the U(1) gauge charge, and which suppress
monopole proliferation for large $N_f$. It was assumed in Ref.~\onlinecite{acl} that $N_f=4$ was large enough for monopole suppression.
However, in the event $N_f > 4$ fermions are required, the holon superconductor would be
unstable to supersolid states, as discussed in the present paper. However, an understanding of the nature of the 
symmetry breaking in these supersolids requires computation of the monopole symmetry properties
in the presence of gapless Dirac fermions --- this we will address in future work.

\acknowledgements
We thank T.~Senthil for very useful discussions, and in particular for clarifying the stability of the doublon metal
against monopole proliferation, and for many valuable comments on the nature of the confining states. 
Our results for the AFM metal to the doublon metal transition have some overlap with
recent results obtained by Senthil \cite{senthilmott} in a different context.
We also thank Tanmoy Das, Pengcheng Dai, Martin Greven, Yong Baek Kim, Sung-Sik Lee, and Vidya Madhavan for valuable discussions.
This research was supported by the NSF under grants
DMR-0132874, DMR-0541988 and DMR-0537077.

\appendix
\section{Time Reversal Symmetry}
\label{app:trs}

Here we outline how the time reversal symmetry was implemented in the symmetry tables. We begin by defining time reversal on the lattice Grassman numbers $c_{\alpha},c^{\alpha\dagger}$:
\begin{eqnarray}
\mathcal{T}[c_{\alpha}]&=& - \varepsilon_{\alpha\beta}c^{\beta\dagger}\\
\mathcal{T}[c^{\alpha\dagger}]&=& \varepsilon^{\alpha\beta}c_{\beta}
\end{eqnarray}
This definition results in: 1) the dynamic term in the action is left invariant under time reversal, 2) the local electron density is invariant under time reversal, and 3) the electronic spin density changes sign under time reversal.

Now using the following transformation for the bosons: $\mathcal{T}[b_{\alpha}]=\varepsilon_{\alpha\beta}b^{\beta\dagger}$ and $\mathcal{T}[\overline{b}^\alpha]=\varepsilon^{\alpha\beta}\overline{b}_{\beta}^\dagger$ (for bosons the conjugates are of course not independent), we can infer what the $g$ should transform into under time reversal [since we know how to write $c$ in terms of $g$ and $b$ and we know how the $c$ transform]. This is recorded in Table~\ref{table0}. The last step is to go from the lattice $f,g$ and $b$ into their continuum counter-parts and this requires knowledge of where the $g$ fields have their minima, but is otherwise straight forward. This is recorded in Table~\ref{psgtable}. Note that unlike the analysis in Ref.~\onlinecite{ffliq} the two sublattice fermions $g_\pm$ transform in the same way. This is related to the position of the doublon pockets in the BZ.

\section{Monopole action in the doublon metal}
\label{sec:monopole}

This appendix will consider the action of a monopole in the U(1) gauge theory of the doublon
metal state of Section~\ref{sec:metal}. The $g_\pm$ Fermi surfaces have low energy excitations
which carry a U(1) gauge charge, and we will discuss their influence on the monopole dynamics.
This problem was originally considered by Herbut {\em et al.} \cite{herbut} using a duality
analysis, but an oversight in their reasoning was pointed out by Hermele {\em et al.} \cite{hermele}.
Here, we will update the analysis of Herbut {\em et al.} and find that the
action of a monopole diverges linearly with system size, consistent with other investigations \cite{hermele,sungsik}.

We begin with the effective action of the U(1) gauge field, $A_\mu$, after the $g_\pm$ fermions and the
$z_\alpha$ have
been integrated out. At quadratic order, this can be written \cite{nagaosa} in terms of the components of the `electromagnetic' field
$F_{\mu\nu} = \partial_\mu A_\nu
- \partial_\nu A_\mu$:
\begin{equation}
\mathcal{S}_{\rm eff} (A) = \frac{1}{2} \int \frac{d^2 k d \omega}{8 \pi^3} \left\{
\varepsilon (k, \omega) F_{i\tau}^2 + \mu (k, \omega) F_{xy}^2 \right\}, \label{sem}
\end{equation}
where $\varepsilon (k, \omega)$ and $\mu (k, \omega)$ are the dielectric constant and the
magnetic permeability respectively. For $|\omega| < v_F k$, in the doublon metal we have \cite{herbut}
\begin{equation}
\mu (k,\omega) \sim 1 + \chi \frac{|\omega|}{k^3} ~~~,~~~\varepsilon (k, \omega) \sim \frac{1}{k^2} \left( 1 + \chi' \frac{|\omega|}{k} \right),
\label{m1}
\end{equation}
while at the O(4) quantum critical point to the AFM metal the critical $z_\alpha$ spinons lead to the propagator in Eq.~(\ref{da})
which corresponds to the magnetic permeability
\begin{equation}
\mu (k,\omega) \sim \frac{1}{k} + \chi \frac{|\omega|}{k^3}.
\label{m2}
\end{equation}
We now apply the duality methods discussed in Section~\ref{sec:dual} to Eq.~(\ref{sem}), and obtain the dual theory
for the `height' field $h$
\begin{equation}
\mathcal{S}_h = \frac{1}{2} \int \frac{d^2 k d \omega}{8 \pi^3} \left\{ \frac{k^2}{\varepsilon (k, \omega)} + \frac{\omega^2}{\mu (k, \omega)}
\right\} h^2 , \label{shh}
\end{equation}
where the monopole operator is $m \sim e^{2 \pi i h}$. Note that for $\omega = 0$, the leading $k$-dependence 
in Eq.~(\ref{shh}) is of order $k^4$. Herbut {\em et al.} \cite{herbut} argued that renormalization effects
would always generate an analytic term of order $k^2$, and proceeded to investigate its consequences.
As noted by Hermele {\em et al.} \cite{hermele}, this is incorrect---the leading term remains $\sim k^4$ 
because it is protected by the presence of the $g_\pm$ Fermi surface.

We can now estimate the action of a monopole in the present Gaussian/RPA approximation \cite{nagaosa}
\begin{equation}
\exp \left( - \mathcal{S}_{m} \right) = \langle m \rangle  = \exp \left( - 2 \pi ^2 \langle h^2 \rangle \right)
\end{equation}
from which we obtain
\begin{equation}
\mathcal{S}_m = 2 \pi^2 \int \frac{d^2 k d \omega}{8 \pi^3} \frac{\mu (k, \omega) \varepsilon (k, \omega)}{k^2 \mu (k, \omega) + \omega^2 \varepsilon (k, \omega)}.
\end{equation}
From Eqs.~(\ref{m1},\ref{m2}) we now find an infrared divergence $\sim \int d^2 k k^{-3}$, which indicates that $\mathcal{S}_m$
diverges linearly with system size. This justifies the neglect of monopoles in the doublon metal and at the O(4) quantum critical point.

\end{document}